\documentclass[%
superscriptaddress,
nofootinbib,
twocolumn,
amsmath,amssymb,
aps,
prd,
a4paper
]{revtex4-2}

\allowdisplaybreaks[1]

\usepackage{newtxtext,newtxmath}
\usepackage{graphicx}
\usepackage{dcolumn}
\usepackage{bm}
\usepackage{color}

\newcommand{\VEC}{\bm}
\newcommand{\MAT}{\bm}

\newcommand{\PP}{\bm{\Psi}}
\newcommand{\YY}{\overline{\bm{\Psi}}}
\newcommand{\SL}{\overline{\bm{s}}}

\begin{document}

\preprint{}

\title{Developing a Theoretical Model for the Resummation of Infrared Effects \\ in the Pre- and Post-Reconstruction Galaxy Bispectra}

\author{Naonori Sugiyama}
\email{nao.s.sugiyama@gmail.com}
\affiliation{National Astronomical Observatory of Japan, Mitaka, Tokyo 181-8588, Japan}
\date{\today}

\begin{abstract}

    While substantial progress has been made in studying the pre-reconstruction galaxy bispectrum, investigations of the post-reconstruction bispectrum are still in nascent stages. In this paper, we present a bispectrum model that incorporates one-loop corrections in the Standard Perturbation Theory (SPT), while simultaneously addressing infrared (IR) effects in the post-reconstruction density fluctuations in a non-perturbative approach. This model accurately captures the nonlinear behavior of the Baryon Acoustic Oscillation (BAO) signal within the post-reconstruction bispectrum framework. Furthermore, the incorporation of the one-loop correction extends its applicability to smaller scales. Given that the approach to addressing IR effects remains invariant before and after the reconstruction process, the resulting IR-resummed bispectrum model exhibits uniformity across both scenarios. Throughout this analysis, we achieve a comprehensive and unified description of the bispectrum that seamlessly spans the pre- and post-reconstruction phases.

\end{abstract}

\maketitle

\section{Introduction}

The three-point correlation function (3PCF) of galaxies, or its corresponding bispectrum in Fourier space, provides additional cosmological information compared to the case only considering the galaxy two-point correlation function (2PCF) and power spectrum. In particular, the isotropic component of the three-point statistics, specifically the \textit{monopole} component of the galaxy 3PCF or bispectrum, has been the primary focus for constraining standard cosmological parameters~\cite{Gil-Marin:2016wya,Slepian:2016kfz,Pearson:2017wtw,DAmico:2019fhj,Philcox:2021kcw} and examining primordial non-Gaussianity~\cite{Cabass:2022wjy,DAmico:2022gki,Cabass:2022ymb}. Recently, by using the anisotropic components of the three-point statistics, namely the \textit{quadrupole} and \textit{hexadecapole} components, more improved constraints on standard cosmological parameters have been achieved~\cite{Sugiyama:2020uil,DAmico:2022osl,Ivanov:2023qzb}. Furthermore, using the quadrupole 3PCF, tests on modified gravity theories and the consistency relation of the Large-Scale Structures (LSS) have been conducted~\cite{Sugiyama:2023tes,Sugiyama:2023zvd}. These data analysis techniques are expected to further evolve in upcoming spectroscopic galaxy surveys such as the Dark Energy Spectroscopic Instrument~\citep[DESI;][]{DESI:2016fyo}\footnote{\url{http://desi.lbl.gov/}}, the Subaru Prime Focus Spectrograph ~\citep[PFS;][]{PFSTeam:2012fqu}\footnote{\url{https://pfs.ipmu.jp/index.html}}, and, Euclid~\citep{EUCLID:2011zbd}\footnote{\url{www.euclid-ec.org}}.

One of the key factors affecting the constraining power of the galaxy bispectrum is the non-Gaussian errors embedded within its covariance matrix. \citet{Sugiyama:2019ike} pointed out that non-Gaussian errors dominate over Gaussian errors in the galaxy bispectrum. This dominance of non-Gaussian errors is a unique statistical characteristic of the bispectrum, significantly different from the power spectrum case, where Gaussian errors are more prevalent. Specifically, the signal-to-noise ratio (S/N) of the bispectrum decreases by a factor of 3 to 4 when non-Gaussian errors are considered, as opposed to considering only Gaussian errors. This finding implies that the presence of non-Gaussian errors significantly limits the constraining power of cosmological parameters derived from the galaxy bispectrum.

The reconstruction of galaxy distributions~\cite{Eisenstein:2006nk} was originally developed to enhance signals from baryon acoustic oscillations~\cite[BAO;][]{Sunyaev:1970eu,Peebles:1970ag}. Recently, it has also been recognized for its ability to minimize non-Gaussian errors in the covariance matrix. Power spectrum analysis has confirmed that this reduction of non-Gaussian errors significantly increases the constraining power for standard cosmological parameters~\cite{Hikage:2020fte,Wang:2022nlx}. This method of error reduction through reconstruction is particularly effective in bispectrum analysis, where \citet{Shirasaki:2020vkk} has shown that post-reconstruction bispectrum analysis can improve constraints on primordial non-Gaussianity by a factor of $3$ to $4$ compared to pre-reconstruction analysis. Given these advancements, a joint analysis of the post-reconstruction power spectrum and bispectrum is anticipated to become the norm in future research.

In the field of cosmology, when analyzing spectroscopic galaxy datasets, such as data from the Baryon Oscillation Spectroscopic Survey \citep[BOSS;][]{Eisenstein:2011sa,Bolton:2012hz,Dawson:2012va,Alam:2015mbd}, we measure both power spectra and bispectra. Subsequently, these measurements are compared with their respective theoretical models. For these analyses, improved perturbation theories, which build upon Standard Perturbation Theory (SPT)~\cite[e.g.,][]{Bernardeau:2001qr}, are frequently employed. The following briefly reviews these improved perturbation theories across four scenarios: pre-reconstruction power spectra, post-reconstruction power spectra, pre-reconstruction bispectra, and post-reconstruction bispectra.

\begin{enumerate}

    \item For pre-reconstruction power spectra, various advanced theoretical models have been proposed and applied in data analysis. These include Convolution Lagrangian Perturbation Theory (CLPT)~\cite{Carlson:2012bu,Wang:2013hwa}, Renormalized Perturbation Theory (RPT)~\cite[][]{Crocce:2005xy}, the TNS model~\cite{Taruya:2010mx}, and the Effective Field Theory of Large-scale Structure (EFT of LSS)~\cite{Baumann:2010tm,Carrasco:2012cv}.

    \item Although there are fewer theoretical studies on pre-reconstruction bispectra compared to those on the power spectrum, models based on EFT~\cite{DAmico:2019fhj,Philcox:2021kcw,DAmico:2022osl,Ivanov:2023qzb} and the resummation of infrared (IR) effects~\cite{Blas:2016sfa,Ivanov:2018gjr,Sugiyama:2020uil,Sugiyama:2023tes,Sugiyama:2023zvd} have been applied to actual galaxy data.

    \item Regarding post-reconstruction power spectra, theoretical calculations using SPT~\cite{Schmittfull:2015mja,Hikage:2017tmm,Hikage:2019ihj} and the Zel'dovich approximation~\cite{Padmanabhan:2008dd,Seo:2015eyw,White:2015eaa,Chen:2019lpf} have been conducted. However, models employed in actual data analyses have predominantly been empirical, characterized by the nonlinear damping of BAO with a single Gaussian damping function. More recently, \citet{Sugiyama:2024eye} and \citet{Chen:2024tfp} have demonstrated that, by appropriately resumming IR effects on post-reconstruction density fluctuations, the nonlinearity of the BAO signal in the post-reconstruction power spectrum can indeed be described with a single Gaussian damping function. This finding provides theoretical support for the validity of previous post-reconstruction BAO analyses.

    \item Theoretical studies on post-reconstruction bispectra remain inadequate. While some studies are based on SPT~\cite{Schmittfull:2015mja,Shirasaki:2020vkk}, no theoretical calculations employing improved perturbation theories beyond SPT have been conducted to date. Furthermore, cosmological analyses using post-reconstruction bispectra have yet to be carried out.

\end{enumerate}

The aim of this paper is to develop a theoretical model for post-reconstruction galaxy bispectra, with the intention of applying it to future cosmological analyses. To this end, we introduce a model that incorporates nonlinear IR effects, marking an initial step towards the development of theoretical models for post-reconstruction bispectra that extend beyond SPT. The emphasis on IR effects is motivated by their unique characteristic of being amenable to non-perturbative treatment. Moreover, it is well-established that these non-perturbative IR effects can accurately describe the nonlinear damping of the BAO signal~\cite{Zeldovich:1969sb,Eisenstein:2006nj,Crocce:2007dt,Matsubara:2007wj}. Therefore, it is essential to properly consider IR effects in reconstruction processes that are sensitive to the BAO signal.

\citet{Sugiyama:2024eye} has shown that IR effects can be represented as a coordinate transformation of density fluctuations via a displacement vector, both before and after reconstruction process. This finding suggests that IR effects can be uniformly addressed, regardless of whether reconstruction has occurred. Consequently, it is feasible to apply the IR-resummed model of the pre-reconstruction bispectrum at the leading order (tree level), initially proposed by~\citet{Sugiyama:2020uil}, to the post-reconstruction bispectrum as well. However, this application necessitates the substitution of certain components, such as nonlinear kernel functions, to suit the post-reconstruction scenario. Moreover, this paper presents the calculation of the IR-resummed model including next-leading order terms, i.e., 1-loop correction terms, which are relevant for scales smaller than those addressed by the tree-level solution.

The calculation of the bispectrum is inherently complex, involving a variety of kernel functions. This complexity is particularly prominent in the resummation of IR effects, as this process entails dealing with infinite mode-coupling integrals. In this paper, we develop a method to simplify the calculation of IR effects on the pre-reconstruction bispectrum as performed by \citet{Sugiyama:2020uil}, and apply this method to compute the 1-loop correction terms.

The outline of this paper is as follows. In Section~\ref{Sec:IR}, we review the methods for handling the IR effects before and after reconstruction; Section~\ref{Sec:Bispectrum} reviews the general properties of the bispectrum. Section~\ref{Sec:Tree} presents the IR-resummed model at the tree-level. Section~\ref{Sec:Oneloop} then derives the main result of this paper, an IR-resummed post-reconstruction bispectrum model incorporating 1-loop corrections. Section~\ref{Sec:Conclusions} serves as the conclusion of this paper. In Appendix~\ref{Sec:Kernel}, we present the kernel functions representing the nonlinear effects for the density fluctuations up to the fourth order, both before and after reconstruction. Appendix~\ref{Sec:Gamma} provides an explanation on the general decomposition methods for the bispectrum using the $\Gamma$-expansion~\cite{Bernardeau:2008fa}. In Appendix~\ref{Sec:Gaussian}, we give the expression for the smoothing factors characterizing the exponential damping function that appears when resumming the IR effects, specifically for the post-reconstruction case. Appendix~\ref{Sec:HighK} gives the solution of the IR (high-k) limit of the 1-loop bispectrum for dark matter in real space.

\section{IR effects on pre- and post-reconstruction density fluctuations}
\label{Sec:IR}

In this section, we briefly review the behavior of density fluctuations in the infrared (IR) limit, before and after reconstruction. Our treatment of the IR limit is based on \citet{Sugiyama:2024eye}, as well as foundational works~\cite{Sugiyama:2013pwa,Sugiyama:2013gza,Sugiyama:2020uil}. For alternative approaches to handling the IR limit, see also~\cite{Baldauf:2015xfa,Vlah:2015zda,Senatore:2014via,Blas:2016sfa,Senatore:2017pbn,Ivanov:2018gjr,Lewandowski:2018ywf}.

\subsection{Pre-reconstruction}

In the Lagrangian description, the displacement vector connecting Eulerian coordinates $\VEC{x}$ to Lagrangian coordinates $\VEC{q}$ is given by 
\begin{eqnarray}
    \VEC{x} = \VEC{q} + \PP(\VEC{q})\;.
\end{eqnarray}
When considering the redshift-space distortion (RSD) effect~\cite{Kaiser:1987qv}, it can be expressed as
\begin{eqnarray}
    \PP_{\rm red}(\VEC{q}) = \PP(\VEC{q}) + \frac{a\dot{\PP}(\VEC{q})\cdot\hat{n}}{aH}\hat{n}\;,
\end{eqnarray}
where $\dot{\PP}(\VEC{q})$ is the time-derivative of $\PP(\VEC{q})$, $\hat{n}$ is the unit vector in the line-of-sight direction, and $a$ and $H$ are the scale factor and the Hubble parameter, respectively. The galaxy density fluctuation is then represented as~\cite{Matsubara:2008wx}
\begin{eqnarray}
    \hspace{-0.3cm}
    1 + \delta_{\rm g}(\VEC{x},\hat{n}) = 
    \hspace{-0.1cm}\int \hspace{-0.15cm} d^3q 
    \left[ 1 + \delta_{\rm b}(\VEC{q}) \right]
    \delta_{\rm D}\left( \VEC{x} - \VEC{q} - \PP_{\rm red}(\VEC{q},\hat{n}) \right)\;,
    \label{Eq:delta_g}
\end{eqnarray}
where $\delta_{\rm b}$ represents the biased density fluctuation in the Lagrangian description, and $\delta_{\rm D}$ denotes the three-dimensional delta function. 

When considering the IR effects, we decompose the displacement vector into its value evaluated at the origin $\VEC{q}=\VEC{0}$ and the other values:
\begin{eqnarray}
    \PP_{\rm red}(\VEC{q},\hat{n}) = \YY_{\rm red}(\hat{n}) + \PP_{\rm (S)red}(\VEC{q},\hat{n})\;,
    \label{Eq:YY}
\end{eqnarray}
where $\YY_{\rm red}(\hat{n})=\PP_{\rm red}(\VEC{q}=\VEC{0},\hat{n})$. Then, $\delta_{\rm g}$ can be formally rewritten as~\cite{Sugiyama:2013gza,Sugiyama:2020uil}
\begin{eqnarray}
    \delta_{\rm g}(\VEC{x},\hat{n}) = \delta_{\rm (S)g}(\VEC{x}-\YY_{\rm red}(\hat{n}),\hat{n})\;,
    \label{Eq:delta_g_2}
\end{eqnarray}
where the short-wavelength density fluctuation $\delta_{\rm (S)g}$ is obtained by replacing $\PP_{\rm red}$ in Eq.~(\ref{Eq:delta_g}) with $\PP_{\rm (S)red}$. In Fourier space, Eq.~(\ref{Eq:delta_g_2}) becomes
\begin{eqnarray}
    \widetilde{\delta}_{\rm g}(\VEC{k},\hat{n})
    = e^{-i\VEC{k}\cdot\YY_{\rm red}(\hat{n})} \widetilde{\delta}_{\rm (S)g}(\VEC{k},\hat{n})\;.
    \label{Eq:delta_g_2_Fourier}
\end{eqnarray}
Throughout this paper, the tilde denotes a Fourier-transformed quantity~\footnote{Our convention for the Fourier transform is
\begin{eqnarray}
    \widetilde{f}(\VEC{k}) = \int d^3x e^{-i\VEC{k}\cdot\VEC{x}} f(\VEC{x})\;.   \nonumber
\end{eqnarray}
}.

In this paper, three assumptions are made when taking the IR limit~\cite{Sugiyama:2024eye}:
\begin{enumerate}
    \item The dominant term in the IR limit arises from the contribution of $\YY_{\rm red}$.
    \item $\delta_{\rm (S)g}$ and $\YY_{\rm red}$ are uncorrelated.
    \item $\delta_{\rm (S)g}$ is truncated at a finite perturbative order, and higher-order perturbative effects than $\delta_{\rm (S)g}$ arise from $\YY_{\rm red}$.
\end{enumerate}

\subsection{Post-reconstruction}
\label{Sec:PostReconstruction}

Using the galaxy density fluctuation, $\delta_{\rm g}(\VEC{x})$, we construct the displacement vector for reconstruction as~\cite{Eisenstein:2006nk}
\begin{eqnarray}
    \hspace{-0.7cm}
    \VEC{s}(\VEC{x},\hat{n}) = 
    \int \frac{d^3p}{(2\pi)^3} e^{i\VEC{p}\cdot\VEC{x}} \left( \frac{i\VEC{p}}{p^2} \right)
    \left(  - \frac{W_{\rm G}(pR_{\rm s})}{b_{1, \rm fid}} \right)
    \widetilde{\delta}_{\rm g}(\VEC{p},\hat{n})\;,
    \label{Eq:S}
\end{eqnarray}
where $b_{1, \rm fid}$ is an input fiducial linear bias parameter for reconstruction, $W_{\rm G}(pR_{\rm s}) = \exp\left( -p^2R_{\rm s}^2/2 \right)$ is a Gaussian filter function, and $R_{\rm s}$ is an input smoothing scale. In the Eulerian description, the density fluctuation after reconstruction is then given by~\cite{Shirasaki:2020vkk,Sugiyama:2020uil}
\begin{eqnarray}
    \delta_{\rm rec}(\VEC{x},\hat{n})
    = \int d^3x' \delta_{\rm g}(\VEC{x}',\hat{n}) 
    \delta_{\rm D}(\VEC{x} - \VEC{x}' - \VEC{s}(\VEC{x}',\hat{n}))\;.
    \label{Eq:delta_rec}
\end{eqnarray}

When taking the IR limit in the post-reconstruction density fluctuation, it is important to consider the IR effects in the pre-reconstruction density fluctuation that constitute $\VEC{s}(\VEC{x})$. By substituting Eq.~(\ref{Eq:delta_g_2_Fourier}) into Eq.~(\ref{Eq:S}), we derive
\begin{eqnarray}
    \VEC{s}(\VEC{x},\hat{n}) = \VEC{s}_{\rm (S)}(\VEC{x}-\YY_{\rm red}(\hat{n}),\hat{n})\;,
\end{eqnarray}
where $\VEC{s}_{\rm (S)}$ is obtained by replacing $\widetilde{\delta}_{\rm g}$ in Eq.~(\ref{Eq:S}) with $\widetilde{\delta}_{\rm (S)g}$. Furthermore, in accordance with Eq.~(\ref{Eq:YY}), we decompose $\VEC{s}_{\rm (S)}$ into its value at the origin and the other values:
\begin{eqnarray}
    \VEC{s}_{\rm (S)}(\VEC{x},\hat{n}) = \SL(\hat{n}) + \VEC{s}_{\rm (SS)}(\VEC{x},\hat{n})\;,
\end{eqnarray}
where $\SL(\hat{n}) = \VEC{s}_{\rm (S)}(\VEC{x}=\VEC{0},\hat{n})$. Eventually, this leads to
\begin{eqnarray}
    \delta_{\rm rec}(\VEC{x},\hat{n}) = \delta_{\rm (S)rec}(\VEC{x}-\YY_{\rm rec}(\hat{n}),\hat{n})\;,
    \label{Eq:delta_rec_2}
\end{eqnarray}
where
\begin{eqnarray}
    \delta_{\rm (S)rec}(\VEC{x},\hat{n})
    &=& \int d^3x' \delta_{\rm (S)g}(\VEC{x}',\hat{n}) \nonumber \\
    &\times&
    \delta_{\rm D}(\VEC{x} - \VEC{x}' - \VEC{s}_{\rm (SS)}(\VEC{x}',\hat{n}))\;.
    \label{Eq:delta_rec_S}
\end{eqnarray}
and
\begin{eqnarray}
    \YY_{\rm rec}(\hat{n}) = \YY_{\rm red}(\hat{n}) + \SL(\hat{n})\;.
\end{eqnarray}
In Fourier space, Eq.~(\ref{Eq:delta_rec_2}) becomes
\begin{eqnarray}
    \widetilde{\delta}_{\rm rec}(\VEC{k},\hat{n}) = e^{-i\VEC{k}\cdot\YY_{\rm rec}(\hat{n})}
    \widetilde{\delta}_{\rm (S)rec}(\VEC{k},\hat{n})\;.
    \label{Eq:delta_rec_2_Fourier}
\end{eqnarray}
Based on Eqs.~(\ref{Eq:delta_rec_2}) and (\ref{Eq:delta_rec_2_Fourier}) above, the IR effect still manifests as a coordinate transformation within the short-wavelength density fluctuation $\delta_{\rm (S)rec}$, even after reconstruction, similar to its behavior before reconstruction. Thus, the treatment of the IR effect remains unchanged before and after reconstruction. Although this paper primarily addresses the post-reconstruction IR effect, it is important to note that out analysis also includes pre-reconstruction scenarios as $R_{\rm s}\to \infty$ approaches infinity.

\subsection{Perturbation expansion}

In cosmological perturbation theory, linear dark matter density fluctuations, denoted as $\delta_{\rm lin}$, serve as the foundation for defining perturbation expansions. Within this framework, the $n$th-order term of a quantity $X$ is expressed as $X^{[n]}$, where $X^{[n]} = {\cal O}([\delta_{\rm lin}]^n)$. In this paper, we use the superscript $[n]$ to denote the $n$th-order term in the perturbation expansion and any associated quantities.

The linear dark matter density fluctuations $\delta_{\rm lin}$ can be decomposed into their time and space dependences and expressed as $\delta_{\rm lin} = D \delta_{\rm lin,0}$. Here, $D$ represents the linear growth factor that characterizes the time dependence, and $\delta_{\rm lin,0}$ is the linear dark matter density fluctuation at $z=0$. The linear growth rate function that characterizes the first-order time derivative of the linear dark matter fluctuations is then defined as $f = d \ln D/d \ln a$.

The $n$th-order post-reconstruction density fluctuation is given by
\begin{eqnarray}
    \widetilde{\delta}_{\rm rec}^{\,[n]}(\VEC{k},\hat{n})
    &=& \int \frac{d^3p_1}{(2\pi)^3}\cdots\int \frac{d^3p_n}{(2\pi)^3}
    (2\pi)^3\delta_{\rm D}\left( \VEC{k} - \VEC{p}_{[1,n]} \right) \nonumber \\
    &\times& Z_{\rm rec}^{[n]}(\VEC{p}_1,\cdots,\VEC{p}_n,\hat{n}) \widetilde{\delta}_{\rm lin}(\VEC{p}_1)\cdots\widetilde{\delta}_{\rm lin}(\VEC{p}_n)\;,
    \label{Eq:Zrec}
\end{eqnarray}
where $\VEC{p}_{[1,n]} = \VEC{p}_1+\cdots+\VEC{p}_n$, and $Z_{\rm rec}^{[n]}$ is the kernel function, including the galaxy bias, RSD, and reconstruction effects. At the first order, the kernel function remains the same before and after reconstruction and is given by~\cite{Kaiser:1987qv}
\begin{eqnarray}
    Z_{\rm rec}^{[1]}(\VEC{p}_1,\hat{n}) = b_1 + f(\hat{p}_1\cdot\hat{n})^2\;, 
\end{eqnarray}
where $b_1$ is the linear bias parameter, and $\hat{p}=\VEC{p}/|\VEC{p}|$. The specific forms of the kernel functions up to the forth order, $Z_{\rm rec}^{[2]}$, $Z_{\rm rec}^{[3]}$, and $Z_{\rm rec}^{[4]}$, are summarized in Appendix~\ref{Sec:Kernel}.

The $n$th-order displacement vector is expressed as
\begin{eqnarray}
    \widetilde{\PP}^{\,[n]}(\VEC{k})
    &=&i\, \int \frac{d^3p_1}{(2\pi)^3}\cdots\int \frac{d^3p_n}{(2\pi)^3}
    (2\pi)^3\delta_{\rm D}\left( \VEC{k} - \VEC{p}_{[1,n]} \right) \nonumber \\
    &\times& \VEC{L}^{[n]}(\VEC{p}_1,\cdots,\VEC{p}_n)\,
    \widetilde{\delta}_{\rm lin}(\VEC{p}_1)\cdots\widetilde{\delta}_{\rm lin}(\VEC{p}_n)\;,
\end{eqnarray}
where $\VEC{L}^{[n]}$ is the $n$th-order kernel function. In redshift space, $\widetilde{\PP}_{\rm red}$ is represented by replacing $\VEC{L}^{[n]}$ with $\VEC{L}_{\rm red}^{[n]}$ given by~\cite{Matsubara:2007wj}
\begin{eqnarray}
    \hspace{-0.3cm}
    \VEC{L}^{[n]}_{\rm red}(\VEC{p}_1,\cdots,\VEC{p}_n,\hat{n})
    = \MAT{R}^{[n]}(\hat{n})\cdot\VEC{L}^{[n]}(\VEC{p}_1,\cdots,\VEC{p}_n)\;.
\end{eqnarray}
Here, the three-dimensional transformation matrix $\MAT{R}^{[n]}$ is
\begin{eqnarray}
    [\MAT{R}^{[n]}(\hat{n})]_{ij} = \MAT{I}_{ij} + n\, f\, \hat{n}_i\hat{n}_j\;,
    \label{Eq:R}
\end{eqnarray}
where $i,j=1,2,3$, and $\VEC{I}$ is the identity matrix. 

As detailed in the subsequent section, we conduct 1-loop bispectrum calculations. The perturbative orders of density fluctuations required for these calculations are up to the fourth order. Therefore, we truncate the short-wavelength density fluctuation at the fourth order. Below, we summarize the relation between $\widetilde{\delta}_{\rm rec}$ and $\widetilde{\delta}_{\rm (S)rec}$ up to the fourth order:
\begin{eqnarray}
    \widetilde{\delta}_{\rm rec}^{\,[1]} &=& \widetilde{\delta}_{\rm (S)rec}^{\,[1]}\;, \nonumber \\
    \widetilde{\delta}_{\rm rec}^{\,[2]} &=& \widetilde{\delta}_{\rm (S)rec}^{\,[2]} + (-i\VEC{k}\cdot\YY^{[1]}_{\rm rec})\widetilde{\delta}^{\,[1]}_{\rm (S)rec}\;,
    \nonumber \\
    \widetilde{\delta}_{\rm rec}^{\,[3]} &=& \widetilde{\delta}_{\rm (S)rec}^{\,[3]} + (-i\VEC{k}\cdot\YY^{[2]}_{\rm rec})\widetilde{\delta}^{\,[1]}_{\rm (S)rec}\nonumber \\
                                       &+& (-i\VEC{k}\cdot\YY^{[1]}_{\rm rec})\widetilde{\delta}^{\,[2]}_{\rm (S)rec} 
                                       +\frac{1}{2}(-i\VEC{k}\cdot\YY^{[1]}_{\rm rec})^2\widetilde{\delta}^{\,[1]}_{\rm (S)rec}\;, \nonumber \\
    \widetilde{\delta}_{\rm rec}^{\,[4]} &=& \widetilde{\delta}_{\rm (S)rec}^{\,[4]} + (-i\VEC{k}\cdot\YY^{[3]}_{\rm rec})\widetilde{\delta}^{\,[1]}_{\rm (S)rec}
    \nonumber \\
                                       &+& (-i\VEC{k}\cdot\YY^{[2]}_{\rm rec})\widetilde{\delta}^{\,[2]}_{\rm (S)rec} 
                                       + (-i\VEC{k}\cdot\YY^{[1]}_{\rm rec})\widetilde{\delta}^{\,[3]}_{\rm (S)rec} \nonumber \\
                                       &+&(-i\VEC{k}\cdot\YY^{[1]}_{\rm rec})(-i\VEC{k}\cdot\YY^{[2]}_{\rm rec})\widetilde{\delta}^{\,[1]}_{\rm (S)rec} \nonumber \\
                                       &+&\frac{1}{2}(-i\VEC{k}\cdot\YY^{[1]}_{\rm rec})^2\widetilde{\delta}^{\,[2]}_{\rm (S)rec} \nonumber \\
                                       &+&\frac{1}{3!}(-i\VEC{k}\cdot\YY^{[1]}_{\rm rec})^3\widetilde{\delta}^{\,[1]}_{\rm (S)rec}\;.
                                       \label{Eq:delta_S_PT}
\end{eqnarray}
Furthermore, in the coordinate transformation that describes the IR effect, we consider only the first-order $\YY_{\rm rec}$. This approach is justified since terms in $\delta_{\rm (S)rec}$ with $\YY_{\rm rec}^{[n\geq2]}$ do not contribute to the 1-loop bispectrum, as demonstrated in the following section. Therefore, the post-reconstruction density fluctuation discussed in this paper can be approximated as follows:
\begin{eqnarray}
    \delta_{\rm rec}(\VEC{k},\hat{n})
    \approx  e^{-i\VEC{k}\cdot\YY^{[1]}_{\rm rec}(\hat{n})}
    \sum_{n=1}^{4}\widetilde{\delta}^{[n]}_{\rm (S)rec}(\VEC{k},\hat{n})\;.
\end{eqnarray}

\section{General properties in the bispectrum}
\label{Sec:Bispectrum}

\subsection{Notational conventions}

To simplify notation, the subscript ``rec'' used to represent reconstructed quantities for density fluctuations and displacement vectors will be omitted throughout the remainder of this paper. Furthermore, we do not explicitly write the dependency on the line-of-sight direction, $\hat{n}$, due to the RSD effect. Thus, we denote $\delta_{\rm rec}(\VEC{x},\hat{n})$ and $\PP_{\rm rec}(\VEC{q},\hat{n})$ as $\delta(\VEC{x})$ and $\PP(\VEC{q})$, respectively.

The 3PCF is defined as the ensemble average of the product of density fluctuations evaluated at three distinct points, $\VEC{x}_1$, $\VEC{x}_2$, and $\VEC{x}_3$. It is characterized by two relative distances, $\VEC{r}_1=\VEC{x}_1-\VEC{x}_3$ and $\VEC{r}_2=\VEC{x}_2-\VEC{x}_3$, due to statistical translational symmetry:
\begin{eqnarray}
    \zeta(\VEC{r}_1,\VEC{r}_2) &=& 
    \langle \delta(\VEC{x}_1) \delta(\VEC{x}_2) \delta(\VEC{x}_3) \rangle \nonumber \\
    &=&\langle \delta(\VEC{x}_1-\VEC{x}_3) \delta(\VEC{x}_2-\VEC{x}_3) \delta(\VEC{0}) \rangle\;.
\end{eqnarray}
In Fourier space, the bispectrum is expressed as
\begin{eqnarray}
    \langle \widetilde{\delta}(\VEC{k}_1) \widetilde{\delta}(\VEC{k}_2) \widetilde{\delta}(\VEC{k}_3) \rangle 
   = (2\pi)^3\delta_{\rm D}(\VEC{k}_{123}) 
   B(\VEC{k}_1,\VEC{k}_2,\VEC{k}_3)\;,
\end{eqnarray}
where $\VEC{k}_{123}=\VEC{k}_1+\VEC{k}_2+\VEC{k}_3$. The condition $\VEC{k}_{123}=\VEC{0}$, through the delta function, reflects the statistical translational symmetry.

In the analytical expressions derived from perturbation theory, when primordial non-Gaussianity is absent, the bispectrum is decomposed into three permutations:
\begin{eqnarray}
    && B(\VEC{k}_1,\VEC{k}_2,\VEC{k}_3) \nonumber \\
    &=& B_{12}(\VEC{k}_1,\VEC{k}_2) + B_{13}(\VEC{k}_1,\VEC{k}_3) + B_{23}(\VEC{k}_2,\VEC{k}_3)\;.
   \label{Eq:B12}
\end{eqnarray}
It is important to note that the exclusion of primordial non-Gaussianity constitutes a fundamental assumption in this study. Throughout this paper, our analysis will be confined to $B_{12}$ exclusively. For simplicity, the subscript ``12'' will be omitted hereafter. Consequently, any term denoted as $B(\VEC{k}_1,\VEC{k}_2)$ henceforth should be interpreted as referring the first term on the right-hand side in Eq.~(\ref{Eq:B12}).

In the framework of perturbation theory, we denote the bispectrum as $B^{[nml]}$, calculated as the product of three density fluctuations of order $n$, $m$, and $l$. For example, a bispectrum composed of two linear fluctuations and one second-order fluctuation is denoted as $B^{[112]}$.

\subsection{$\Gamma$-expansion}

\begin{figure*}
    \centering
    \scalebox{0.95}{\includegraphics[width=\textwidth]{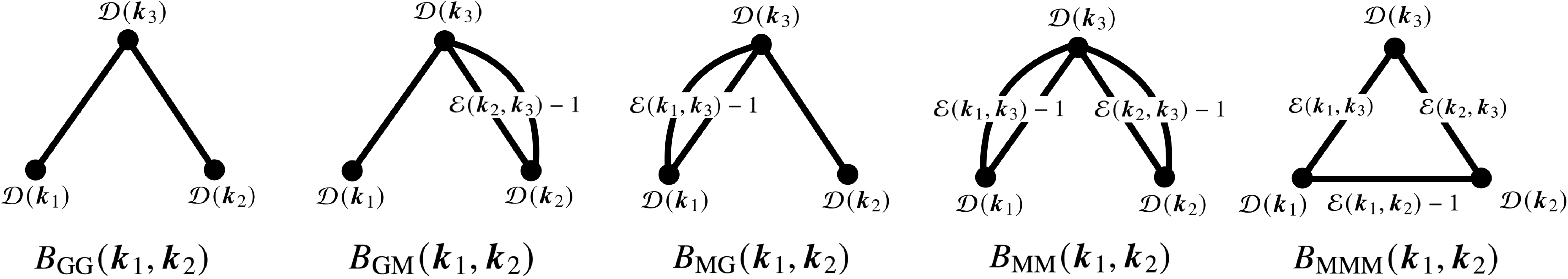}}
    \caption{
Schematic diagram representing the bispectrum components decomposed by the $G$-expansion method as given in Eq.~(\ref{Eq:Bdecom}). Among the lines connecting the vertices corresponding to $\VEC{k}_1$, $\VEC{k}_2$, and $\VEC{k}_3$, closed loops indicate the presence of mode coupling integrals. Note that although only one loop is illustrated, it implicitly shows an infinite number of mode coupling integrals. In the IR limit, each vertex and the lines connecting these vertices are associated with the functions ${\cal D}$ and ${\cal E}$ as defined in Eq.~(\ref{Eq:D_E}). This association enables a straightforward derivation of the non-perturbative solution of each term in the IR limit as given by Eq.~(\ref{Eq:Bdecom_tree}).
    }
    \label{fig:diagrams}
\end{figure*}

The $\Gamma$-expansion~(e.g., \cite{Bernardeau:2008fa}) is a decomposition method that relies on mode-coupling integrals. Using the $\Gamma$-expansion allows for the decomposition of the bispectrum into five components~\cite{Sugiyama:2020uil}, expressed as follows:
\begin{eqnarray}
    B(\VEC{k}_1,\VEC{k}_2)
    &=& B_{\rm GG}(\VEC{k}_1,\VEC{k}_2) P_{\rm lin}(k_1)P_{\rm lin}(k_2) \nonumber \\
    &+& B_{\rm GM}(\VEC{k}_1,\VEC{k}_2) P_{\rm lin}(k_1) \nonumber \\
    &+& B_{\rm MG}(\VEC{k}_1,\VEC{k}_2) P_{\rm lin}(k_2)\nonumber \\
    &+& B_{\rm MM}(\VEC{k}_1,\VEC{k}_2) \nonumber \\
    &+& B_{\rm MMM}(\VEC{k}_1,\VEC{k}_2)\;,
    \label{Eq:Bdecom}
\end{eqnarray}
where $P_{\rm lin}$ denotes the linear dark matter power spectrum, calculated as
\begin{eqnarray}
    \langle \widetilde{\delta}_{\rm lin}(\VEC{k})\widetilde{\delta}_{\rm lin}(\VEC{k}') \rangle
    =(2\pi)^3\delta_{\rm D}(\VEC{k}+\VEC{k}') P_{\rm lin}(k)\;.
\end{eqnarray}
In Eq.~(\ref{Eq:Bdecom}), $B_{\rm GG}$ does not incorporate mode coupling integrals, whereas both $B_{\rm GM}$ and $B_{\rm MG}$ partially include mode coupling integrals. Conversely, $B_{\rm MM}$ and $B_{\rm MMM}$ are composed only of mode-coupling integrals. For detailed expressions of $B_{\rm GG}$, $B_{\rm GM}$, $B_{\rm MG}$, $B_{\rm MM}$, and $B_{\rm MMM}$, see Appendix~\ref{Sec:Gamma}.

Figure~\ref{fig:diagrams} shows a schematic diagram representing the bispectrum components decomposed by the $\Gamma$-expansion method. The bispectrum is represented by three vertices corresponding to $\VEC{k}_1$, $\VEC{k}_2$, and $\VEC{k}_3$, and it can be decomposed into components based on how these wave vectors are connected. Regarding $B_{\rm GG}$, $\VEC{k}_1$ and $\VEC{k}_3$, as well as $\VEC{k}_2$ and $\VEC{k}_3$, are connected without the mode coupling integral. As shown in Section~\ref{Sec:Tree}, the tree-level bispectrum is classified under this term. The $B_{\rm GM}$, $B_{\rm MG}$, and $B_{\rm MM}$ terms respectively involve $\VEC{k}_1$ and $\VEC{k}_3$, $\VEC{k}_2$ and $\VEC{k}_3$, or both being connected through the mode coupling integral. The mode coupling integral is here represented as a closed loop. For simplicity, only one loop is illustrated in the figure, but it is implied that an infinite number of mode coupling integrals are considered. The $B_{\rm MMM}$ term connects $\VEC{k}_1$ and $\VEC{k}_2$ to form a closed triangle diagram. It is noted that the lines represented as straight lines in the $B_{\rm MMM}$ diagram include an infinite number of mode coupling integrals implicitly. For the non-perturbative behavior in the IR limit of each term, see Sections~\ref{Sec:TreeLevelIRcancel} and \ref{Sec:OneLoopIRcancel}.

\subsection{IR cancellation}
\label{Sec:IRcancel}

If $\YY$ and $\delta_{\rm (S)}$ are completely uncorrelated, all IR effects induced by $\YY$ are canceled out due to statistical translational symmetry in the 3PCF calculation, yielding contributions only from short-wavelength density fluctuations:
\begin{eqnarray}
    &&\langle \delta(\VEC{x}_1)\delta(\VEC{x}_2)\delta(\VEC{x}_3)\rangle \nonumber \\
  &=&\langle \delta_{\rm (S)}(\VEC{x}_1-\YY)\delta_{\rm (S)}(\VEC{x}_2-\YY)\delta_{\rm (S)}(\VEC{x}_3-\YY)\rangle
  \nonumber \\
  &=&\langle \delta_{\rm (S)}(\VEC{x}_1)\delta_{\rm (S)}(\VEC{x}_2)\delta_{\rm (S)}(\VEC{x}_3)\rangle\;.
\end{eqnarray}
In Fourier space, this relation can be expressed as
\begin{eqnarray}
    &&\big\langle \widetilde{\delta}(\VEC{k}_1)
    \widetilde{\delta}(\VEC{k}_2)\widetilde{\delta}(\VEC{k}_3)\big\rangle \nonumber \\
  &=&
  \big\langle e^{-i(\VEC{k}_1+\VEC{k}_2+\VEC{k}_3)\cdot\YY}\big\rangle
  \big\langle \widetilde{\delta}_{\rm (S)}(\VEC{k}_1)
    \widetilde{\delta}_{\rm (S)}(\VEC{k}_2)\widetilde{\delta}_{\rm (S)}(\VEC{k}_3)\big\rangle \nonumber \\
    &=&   \big\langle \widetilde{\delta}_{\rm (S)}(\VEC{k}_1)
    \widetilde{\delta}_{\rm (S)}(\VEC{k}_2)\widetilde{\delta}_{\rm (S)}(\VEC{k}_3)\big\rangle \;.
    \label{Eq:IRcancel_B}
\end{eqnarray}
The bispectrum resulting from short-wavelength density fluctuations is represented as
 \begin{eqnarray}
    && \langle \widetilde{\delta}_{\rm (S)}(\VEC{k}_1) 
     \widetilde{\delta}_{\rm (S)}(\VEC{k}_2) 
     \widetilde{\delta}_{\rm (S)}(\VEC{k}_3) \rangle \nonumber \\
   &=& (2\pi)^3\delta_{\rm D}(\VEC{k}_{123})\,
   [B_{\rm (S)}(\VEC{k}_1,\VEC{k}_2)+ \mbox{2 perms.}]\;,
\end{eqnarray}
and is denoted using the subscript ${\rm (S)}$. 

\section{Tree level bispectra}
\label{Sec:Tree}

\subsection{SPT}

The leading-order term, known as the tree-level, is represented by
\begin{eqnarray}
    B_{\rm tree}(\VEC{k}_1,\VEC{k}_2) = B_{\rm GG}^{[112]}(\VEC{k}_1,\VEC{k}_2) P_{\rm lin}(k_1)P_{\rm lin}(k_2)\;,
    \label{Eq:B_tree}
\end{eqnarray}
where
\begin{eqnarray}
    B_{\rm GG}^{[112]}(\VEC{k}_1,\VEC{k}_2) = 2\, Z_{\rm rec}^{[2]}(\VEC{k}_1,\VEC{k}_2) Z_{\rm rec}^{[1]}(\VEC{k}_1) Z_{\rm rec}^{[1]}(\VEC{k}_2)\;.
\end{eqnarray}
Consequently, the bispectrum at the tree level consists only of the $B_{\rm GG}$ term.

Assuming the lack of correlation between $\delta_{\rm (S)}$ and $\YY$ in Eq.~(\ref{Eq:delta_S_PT}), we can show that the tree-level bispectrum is the same as the short-wavelength one:
\begin{eqnarray}
    B_{\rm (S)GG}^{[112]}(\VEC{k}_1,\VEC{k}_2) &=& B_{\rm GG}^{[112]}(\VEC{k}_1,\VEC{k}_2)\;. 
    \label{Eq:B_S_tree}
\end{eqnarray}
This is an expected result from the general discussion in Eq.~(\ref{Eq:IRcancel_B}).

\subsection{IR cancellation}
\label{Sec:TreeLevelIRcancel}

In Fourier space, the IR effect manifests as the exponential function in Eq.~(\ref{Eq:IRcancel_B}). Assuming $\YY\simeq\YY^{[1]}$, it can be represented as
\begin{eqnarray}
    \hspace{-0.5cm}
    \big\langle e^{-i(\VEC{k}_1+\VEC{k}_2+\VEC{k}_3)\cdot\YY^{[1]}}\big\rangle
    &=& {\cal D}(\VEC{k}_1){\cal D}(\VEC{k}_2){\cal D}(\VEC{k}_3) \nonumber \\
    &\times&{\cal E}(\VEC{k}_1,\VEC{k}_2)
    {\cal E}(\VEC{k}_2,\VEC{k}_3)
    {\cal E}(\VEC{k}_1,\VEC{k}_3)\;.
    \label{Eq:e_DE}
\end{eqnarray}
Here, ${\cal D}(\VEC{k})$ and ${\cal E}(\VEC{k},\VEC{k}')$ are defined as follows~\footnote{The cumulants of a statistical quantity $X$, denoted $\langle X^n \rangle_{\rm c}$, are obtained from a power series expansion of the cumulant-generating function $\ln \langle e^{X}\rangle$:
\begin{eqnarray}
    \ln\left(  \langle e^X \rangle \right)
    =  \sum_{n=1}\frac{1}{n!} \langle X^n \rangle_{\rm c}\;.
\end{eqnarray}
}:
\begin{eqnarray}
    {\cal D}(\VEC{k})
   &=&
   \exp\left( \frac{1}{2}\left\langle \left( -i\VEC{k}\cdot\YY^{[1]} \right)^2\right\rangle_{\rm c} \right)\;,
   \nonumber \\
    {\cal E}(\VEC{k}_1,\VEC{k}_2)
   &=&
   \exp\left( \left\langle \left( -i\VEC{k}_1\cdot\YY^{[1]}\right)\left( -i\VEC{k}_2\cdot\YY^{[1]} \right)\right\rangle_{\rm c} \right)\;.
   \label{Eq:D_E}
\end{eqnarray}
The function ${\cal D}(\VEC{k})$ is presented as a two-dimensional exponential damping function, specifically calculated as
\begin{eqnarray}
    {\cal D}(\VEC{k}) = \exp\left( -\frac{ k^2(1-\mu^2)\sigma_{\perp}^2 + k^2\mu^2 \sigma^2_{\parallel} }{2} \right)\;,
    \label{Eq:Damping}
\end{eqnarray}
where $\mu=\hat{n}\cdot\hat{k}$. The radial and transverse components of the smoothing factors, $\sigma_{\perp}^2$ and $\sigma_{\parallel}^2$, are given by Eqs.~(\ref{Eq:Sigma}) and (\ref{Eq:ps_ss}). Furthermore, the function ${\cal E}(\VEC{k}_1,\VEC{k}_2)$ satisfies the following relation:
\begin{eqnarray}
    {\cal E}(\VEC{k}_1,\VEC{k}_2)
    = {\cal D}^{-1}(\VEC{k}_1){\cal D}^{-1}(\VEC{k}_2){\cal D}(\VEC{k}_3) 
    \label{Eq:EE}
\end{eqnarray}
under the condition $\VEC{k}_1+\VEC{k}_2+\VEC{k}_3=\VEC{0}$.

The function ${\cal E}(\VEC{k},\VEC{k}')$ connects two different wave vectors, $\VEC{k}$ and $\VEC{k}'$. In other words, this operation produces mode coupling integrals. On the other hand, the function ${\cal D}(\VEC{k})$ is only associated only with the corresponding wave vector $\VEC{k}$. Figure~\ref{fig:diagrams} shows how the functions ${\cal D}$ and ${\cal E}$ appear in each term of the bispectrum decomposed by the $\Gamma$-expansion in the IR limit. Taking this into account and substituting Eq.~(\ref{Eq:B_S_tree}) and Eq.~(\ref{Eq:e_DE}) into Eq.~(\ref{Eq:IRcancel_B}), we derive
\begin{eqnarray}
     B_{\rm GG,\, IR}(\VEC{k}_1,\VEC{k}_2) 
    &=&{\cal D}(\VEC{k}_1){\cal D}(\VEC{k}_2){\cal D}(\VEC{k}_3) B_{\rm GG}^{[112]}(\VEC{k}_1,\VEC{k}_2) \;, 
    \nonumber \\
     B_{\rm GM,\, IR}(\VEC{k}_1,\VEC{k}_2) 
    &=&{\cal D}(\VEC{k}_1){\cal D}(\VEC{k}_2){\cal D}(\VEC{k}_3) 
    \left(  {\cal E}(\VEC{k}_2,\VEC{k}_3) - 1\right) \nonumber \\
    &\times&
     B_{\rm GG}^{[112]}(\VEC{k}_1,\VEC{k}_2) P_{\rm lin}(k_2)\;, \nonumber \\
    B_{\rm MG,\, IR}(\VEC{k}_1,\VEC{k}_2) 
   &=&{\cal D}(\VEC{k}_1){\cal D}(\VEC{k}_2){\cal D}(\VEC{k}_3) 
   \left(  {\cal E}(\VEC{k}_1,\VEC{k}_3) - 1\right) \nonumber \\
   &\times& B_{\rm GG}^{[112]}(\VEC{k}_1,\VEC{k}_2)P_{\rm lin}(k_1)\;,
    \nonumber \\
     B_{\rm MM,\, IR}(\VEC{k}_1,\VEC{k}_2) 
    &=&{\cal D}(\VEC{k}_1){\cal D}(\VEC{k}_2){\cal D}(\VEC{k}_3)  \nonumber \\
    &\times&
    \left(  {\cal E}(\VEC{k}_2,\VEC{k}_3) - 1\right)
    \left({\cal E}(\VEC{k}_1,\VEC{k}_3) - 1  \right) \nonumber \\
    &\times&
     B_{\rm GG}^{[112]}(\VEC{k}_1,\VEC{k}_2) 
    P_{\rm lin}(k_1) P_{\rm lin}(k_2)\;,
     \nonumber \\
    B_{\rm MMM,\, IR}(\VEC{k}_1,\VEC{k}_2) 
    &=& {\cal D}(\VEC{k}_1){\cal D}(\VEC{k}_2){\cal D}(\VEC{k}_3) 
     \left(  {\cal E}(\VEC{k}_1,\VEC{k}_2) - 1 \right) \nonumber \\
     &\times&
    {\cal E}(\VEC{k}_2,\VEC{k}_3)
    {\cal E}(\VEC{k}_1,\VEC{k}_3) \nonumber \\
    &\times&
    B_{\rm GG}^{[112]}(\VEC{k}_1,\VEC{k}_2)  P_{\rm lin}(k_1) P_{\rm lin}(k_2)\;,
    \label{Eq:B_tree_IR}
\end{eqnarray}
where the subscript ``IR'' indicates a quantity in the IR limit. Furthermore, using Eq.~(\ref{Eq:EE}), we obtain 
\begin{eqnarray}
     B_{\rm GG,\, IR}(\VEC{k}_1,\VEC{k}_2) 
    &=&{\cal D}(\VEC{k}_1){\cal D}(\VEC{k}_2){\cal D}(\VEC{k}_3) B_{\rm GG}^{[112]}(\VEC{k}_1,\VEC{k}_2) \;, 
    \nonumber \\
     B_{\rm GM,\, IR}(\VEC{k}_1,\VEC{k}_2) 
     &=& \big[{\cal D}^2(\VEC{k}_1) - {\cal D}(\VEC{k}_1){\cal D}(\VEC{k}_2){\cal D}(\VEC{k}_3) \big] \nonumber \\
    &\times&
     B_{\rm GG}^{[112]}(\VEC{k}_1,\VEC{k}_2) P_{\rm lin}(k_2)\;, \nonumber \\
    B_{\rm MG,\, IR}(\VEC{k}_1,\VEC{k}_2) 
   &=&
    \big[{\cal D}^2(\VEC{k}_2) - {\cal D}(\VEC{k}_1){\cal D}(\VEC{k}_2){\cal D}(\VEC{k}_3) \big] \nonumber \\
   &\times& B_{\rm GG}^{[112]}(\VEC{k}_1,\VEC{k}_2)P_{\rm lin}(k_1)\;,
    \nonumber \\
     B_{\rm MM,\, IR}(\VEC{k}_1,\VEC{k}_2) 
    &=&
    \big[
        {\cal D}(\VEC{k}_1){\cal D}(\VEC{k}_2){\cal D}^{-1}(\VEC{k}_3) - {\cal D}^2(\VEC{k}_1) \nonumber \\
        && - {\cal D}^2(\VEC{k}_2) + {\cal D}(\VEC{k}_1){\cal D}(\VEC{k}_2){\cal D}(\VEC{k}_3)
    \big] \nonumber \\
    &\times&
     B_{\rm GG}^{[112]}(\VEC{k}_1,\VEC{k}_2) 
    P_{\rm lin}(k_1) P_{\rm lin}(k_2)\;,
     \nonumber \\
    B_{\rm MMM,\, IR}(\VEC{k}_1,\VEC{k}_2) 
    &=&  \big[ 1 - {\cal D}(\VEC{k}_1){\cal D}(\VEC{k}_2){\cal D}^{-1}(\VEC{k}_3) \big] \nonumber \\
    &\times&
    B_{\rm GG}^{[112]}(\VEC{k}_1,\VEC{k}_2)  P_{\rm lin}(k_1) P_{\rm lin}(k_2)\;.
    \label{Eq:Bdecom_tree}
\end{eqnarray}
Consequently, for all the terms in Eq.~(\ref{Eq:Bdecom_tree}), the infinite perturbative orders that emerge due to infrared (IR) effects, i.e., non-perturbative effects, can be computed using the Gaussian damping function ${\cal D}$. Nonetheless, the substitution of all these terms into Eq.~(\ref{Eq:Bdecom}) results in the cancellation of all contributions from the IR effects, thereby isolating the standard tree-level bispectrum.
\begin{eqnarray}
    B_{\rm IR}(\VEC{k}_1,\VEC{k}_2)
    &=& B_{\rm GG,\,IR}(\VEC{k}_1,\VEC{k}_2) P_{\rm lin}(k_1)P_{\rm lin}(k_2) \nonumber \\
    &+& B_{\rm GM,\,IR}(\VEC{k}_1,\VEC{k}_2) P_{\rm lin}(k_1) \nonumber \\
    &+& B_{\rm MG,\,IR}(\VEC{k}_1,\VEC{k}_2) P_{\rm lin}(k_2)\nonumber \\
    &+& B_{\rm MM,\,IR}(\VEC{k}_1,\VEC{k}_2) \nonumber \\
    &+& B_{\rm MMM,\,IR}(\VEC{k}_1,\VEC{k}_2) \nonumber \\
    &=& B_{\rm tree}(\VEC{k}_1,\VEC{k}_2)\;.
\end{eqnarray}
This result serves as a specific example of tree-level perturbative calculations for IR cancellation discussed in the general case in Section~\ref{Sec:IRcancel}.

Eq.~(\ref{Eq:Bdecom_tree}) exhibits the same form as that derived for the pre-reconstruction bispectrum, as detailed in Eqs.~(52)-(56) in \citet{Sugiyama:2020uil}. This similarity arises because the IR effect for the post-reconstruction case can be addressed in a manner analogous to the pre-reconstruction case, as discussed in Section~\ref{Sec:PostReconstruction}.

From Eq.~(\ref{Eq:Bdecom_tree}), it becomes clear that the two terms consisting only of mode-coupling integrals, $B_{\rm MM}$ and $B_{\rm MMM}$, tend to diverge in the high-$k$ limit due to the presence of the ${\cal D}(\VEC{k}_1){\cal D}(\VEC{k}_2){\cal D}^{-1}(\VEC{k}_3)$ term. Nevertheless, by considering the combination of $B_{\rm MM}$ and $B_{\rm MMM}$, i.e., $B_{\rm MM}+B_{\rm MMM}$, the ${\cal D}(\VEC{k}_1){\cal D}(\VEC{k}_2){\cal D}^{-1}(\VEC{k}_3)$ term cancels out, resulting in a finite value at small scales, as demonstrated in Figure $3$ in \citet{Sugiyama:2020uil}.

\subsection{IR-resummed model}
\label{Sec:IR_resummed_model}

It is noteworthy that among terms related to mode-coupling integrals, such as $B_{\rm GM}$, $B_{\rm MG}$, $B_{\rm MM}$, and $B_{\rm MMM}$, in Eq.~(\ref{Eq:Bdecom_tree}), the linear power spectrum $P_{\rm lin}$ containing the BAO signal emerges. However, the mode-coupling integral is expected to smooth out the BAO signal in $P_{\rm lin}$. This expectation indicates that the assumption that $\delta_{\rm (S)}$ and $\YY$ being uncorrelated does not fully hold, leading to incomplete IR cancellation due to the mode coupling effect.

To account for smoothing effect of the BAO signal, we decompose the linear matter power spectrum into two components: the ``wiggle'' component with only the BAO signal, and the ``no-wiggle'' component lacking the BAO signal. This decomposition is expressed as $P_{\rm lin} = P_{\rm w} + P_{\rm nw}$~\cite{Eisenstein:1997ik,Hamann:2010pw,Chudaykin:2020aoj}, where the subscripts ``w'' and ``nw'' denote ``wiggle'' and ``no-wiggle'', respectively. Subsequently, we replace all $P_{\rm lin}$ in Eqs.~(\ref{Eq:Bdecom_tree}) with $P_{\rm nw}$.

Substituting Eq.~(\ref{Eq:Bdecom_tree}) into Eq.~(\ref{Eq:Bdecom}) after the above replacement between $P_{\rm lin}$ and $P_{\rm nw}$, we derive the post-reconstruction tree-level bispectrum with IR resummation as
\begin{eqnarray}
    && B(\VEC{k}_1,\VEC{k}_2) \nonumber \\
    &=&
    B_{\rm GG}^{[112]}(\VEC{k}_1,\VEC{k}_2) 
    {\cal D}(\VEC{k}_1){\cal D}(\VEC{k}_2){\cal D}(\VEC{k}_3)P_{\rm w}(k_1)P_{\rm w}(k_2) \nonumber\\
    &+&
    B_{\rm GG}^{[112]}(\VEC{k}_1,\VEC{k}_2) {\cal D}^2(\VEC{k}_1) P_{\rm w}(k_1)P_{\rm nw}(k_2) \nonumber\\
    &+&
     B_{\rm GG}^{[112]}(\VEC{k}_1,\VEC{k}_2){\cal D}^2(\VEC{k}_2)P_{\rm nw}(k_1)P_{\rm w}(k_2) \nonumber\\
    &+& 
     B_{\rm tree,\, nw}(\VEC{k}_1,\VEC{k}_2) \;,
    \label{Eq:main1}
\end{eqnarray}
where the no-wiggle version of the tree-level bispectrum is represented by
\begin{eqnarray}
    B_{\rm tree,\, nw}(\VEC{k}_1,\VEC{k}_2)
    = B_{\rm GG}^{[112]}(\VEC{k}_1,\VEC{k}_2)P_{\rm nw}(k_1)P_{\rm nw}(k_2) \;.
    \label{Eq:B_tree_nw}
\end{eqnarray}
This is the first main result of this paper. In our model, the nonlinearity of the BAO signal can be divided into three distinct components: ${\cal D}(\VEC{k}_1){\cal D}(\VEC{k}_2){\cal D}(\VEC{k}_3)P_{\rm w}(k_1)P_{\rm w}(k_2)$, ${\cal D}^2(\VEC{k}_1)P_{\rm w}(k_1)P_{\rm nw}(k_2)$, and ${\cal D}^2(\VEC{k}_2)P_{\rm nw}(k_1)P_{\rm w}(k_2)$, which serve to dampen the BAO signal. The shape of the bispectrum is represented through the no-wiggle version of the tree-level solution in Eq.~(\ref{Eq:B_tree_nw}).

Eq.~(\ref{Eq:main1}) is consistent with the pre-reconstruction bispectrum model with IR resummation, as detailed in Eq.~(1) of \citet{Sugiyama:2020uil}. Note that the post-reconstruction case requires the use of the post-reconstruction kernel functions in Eq.~(\ref{Eq:PostReconZ}) and the post-reconstruction Gaussian smoothing parameters in Eq.~(\ref{Eq:Sigma}).

It is noteworthy that the derivation of Eq.~(\ref{Eq:main1}) benefits from the definition of the function ${\cal E}$ in Eq.~(\ref{Eq:D_E}) during the process. This approach allows us to avoid the complex calculations associated with kernel functions, a methodology previously employed by \citet{Sugiyama:2020uil}. Consequently, this simplification significantly streamlines the computational process, facilitating the straightforward inclusion of additional 1-loop correction terms in the subsequent section.

\section{1-loop level bispectra}
\label{Sec:Oneloop}

\subsection{SPT}

The next-leading order term in SPT, the 1-loop correction, can be decomposed into the following five components:
\begin{eqnarray}
    && B_{\rm 1\mathchar`-loop}(\VEC{k}_1,\VEC{k}_2) \nonumber \\
    &=& \left[B_{\rm GG}^{[114]}(\VEC{k}_1,\VEC{k}_2) + B_{\rm GG}^{\rm [123-I]}(\VEC{k}_1,\VEC{k}_2)  \right]P_{\rm lin}(k_1)P_{\rm lin}(k_2) \nonumber \\
    &+& B^{\rm [123-II]}_{\rm GM}(\VEC{k}_1,\VEC{k}_2) P_{\rm lin}(k_1) + B^{\rm [123-III]}_{\rm MG}(\VEC{k}_1,\VEC{k}_2) P_{\rm lin}(k_2)\nonumber \\
    &+& B_{\rm MMM}^{[222]}(\VEC{k}_1,\VEC{k}_2) \;,
    \label{Eq:B_1loop}
\end{eqnarray}
where
\begin{widetext}
\begin{eqnarray}
    B_{\rm GG}^{[114]}(\VEC{k}_1,\VEC{k}_2)
    &=&  12\,
    Z_{\rm rec}^{[1]}(\VEC{k}_1) 
    Z_{\rm rec}^{[1]}(\VEC{k}_2)
    \int \frac{d^3p}{(2\pi)^3}
    Z_{\rm rec}^{[4]}(\VEC{k}_1,\VEC{k}_2,\VEC{p},-\VEC{p})P_{\rm lin}(p)\;,\nonumber \\
    B_{\rm GG}^{\rm [123-I]}(\VEC{k}_1,\VEC{k}_2)
    &=& 6\,  Z_{\rm rec}^{[2]}(\VEC{k}_1,\VEC{k}_2) 
    \int \frac{d^3p}{(2\pi)^3} 
    \left[ Z_{\rm rec}^{[1]}(\VEC{k}_1)Z_{\rm rec}^{[3]}(\VEC{k}_2,\VEC{p},-\VEC{p}) 
    + Z_{\rm rec}^{[1]}(\VEC{k}_2)Z_{\rm rec}^{[3]}(\VEC{k}_1,\VEC{p},-\VEC{p}) \right]P_{\rm lin}(p)\;,
    \nonumber \\
    B_{\rm GM}^{\rm [123-II]}(\VEC{k}_1,\VEC{k}_2)
    &=& 6\, Z_{\rm rec}^{[1]}(\VEC{k}_1) 
    \int \frac{d^3p}{(2\pi)^3} Z_{\rm rec}^{[2]}(\VEC{p},\VEC{k}_2-\VEC{p})Z_{\rm rec}^{[3]}(\VEC{k}_1,\VEC{p},\VEC{k}_2-\VEC{p}) P_{\rm lin}(p) P_{\rm lin}(|\VEC{k}_2-\VEC{p}|)\;,
    \nonumber \\
    B_{\rm MG}^{\rm [123-III]}(\VEC{k}_1,\VEC{k}_2)
    &=& 6\, Z_{\rm rec}^{[1]}(\VEC{k}_2)  
    \int \frac{d^3p}{(2\pi)^3} Z_{\rm rec}^{[2]}(\VEC{p},\VEC{k}_1-\VEC{p})Z_{\rm rec}^{[3]}(\VEC{k}_2,\VEC{p},\VEC{k}_1-\VEC{p}) P_{\rm lin}(p) P_{\rm lin}(|\VEC{k}_1-\VEC{p}|)\;,
    \nonumber \\
    B_{\rm MMM}^{[222]}(\VEC{k}_1,\VEC{k}_2) 
    &=&  
    \frac{8}{3}\,\int \frac{d^3p}{(2\pi)^3}
    Z_{\rm rec}^{[2]}(\VEC{p},\VEC{k}_1-\VEC{p}) 
    Z_{\rm rec}^{[2]}(-\VEC{p},\VEC{k}_2+\VEC{p})
    Z_{\rm rec}^{[2]}(\VEC{k}_2+\VEC{p},\VEC{k}_1-\VEC{p})
    P_{\rm lin}(p) P_{\rm lin}(|\VEC{k}_2+\VEC{p}|)P_{\rm lin}(|\VEC{k}_1-\VEC{p}|)\;. \nonumber \\
    \label{Eq:B1loop}
\end{eqnarray}
\end{widetext}
Here, it should be noted that the $B_{\rm MM}$ term arises from 2-loop and higher-order corrections, and therefore, does not appear in the 1-loop corrections.

The IR (high-$k$) limit solutions for the 1-loop bispectrum corrections are obtained by substituting the post-reconstruction kernel functions in the IR limit, as provided by Eq.~(\ref{Eq:Z_rec_IR}), into Eq.~(\ref{Eq:B1loop}). It is important to note that the conditions for the IR limit in the bispectrum vary with each term of the bispectrum: for $B_{\rm GG}^{[114]}$ and $B_{\rm GG}^{\rm [123-I]}$, it is $p\to0$; for $B_{\rm GM}^{\rm [123-II]}$, it is $p\to0$ and $|\VEC{k}_2-\VEC{p}|\to0$; $B_{\rm MG}^{\rm [123-III]}$, it is $p\to0$ and $|\VEC{k}_1-\VEC{p}|\to0$; for $B_{\rm MMM}^{[222]}$, it is $p\to0$, $|\VEC{k}_1-\VEC{p}|\to 0$, and $|\VEC{k}_2+\VEC{p}|\to0$. Then, we obtain
\begin{eqnarray}
    B_{\rm GG,\,IR}^{[114]}(\VEC{k}_1,\VEC{k}_2) &=&  
    \left[ \ln {\cal D}(\VEC{k}_3) \right]
    B_{\rm GG}^{[112]}(\VEC{k}_1,\VEC{k}_2)\;, \nonumber \\
    B_{\rm GG,\,IR}^{\rm [123-I]}(\VEC{k}_1,\VEC{k}_2) 
    &=&  
    \left[ \ln \left(  {\cal D}(\VEC{k}_1){\cal D}(\VEC{k}_2) \right)\right]
    B_{\rm GG}^{[112]}(\VEC{k}_1,\VEC{k}_2)\;, \nonumber \\
    B_{\rm GM,\,IR}^{\rm [123-II]}(\VEC{k}_1,\VEC{k}_2) &=&  
    \left[ \ln \left(  {\cal D}(\VEC{k}_1){\cal D}^{-1}(\VEC{k}_2){\cal D}^{-1}(\VEC{k}_3)\right)   \right] 
    \nonumber \\
    &\times& 
    B_{\rm GG}^{[112]}(\VEC{k}_1,\VEC{k}_2) P_{\rm lin}(k_2)\;, \nonumber \\
    B_{\rm MG,\,IR}^{\rm [123-III]}(\VEC{k}_1,\VEC{k}_2) &=&  
    \left[  \ln \left( {\cal D}(\VEC{k}_2){\cal D}^{-1}(\VEC{k}_1){\cal D}^{-1}(\VEC{k}_3)  \right) \right]
    \nonumber \\
    &\times&
    B_{\rm GG}^{[112]}(\VEC{k}_1,\VEC{k}_2)P_{\rm lin}(k_1)\;, \nonumber \\
    B_{\rm MMM,\,IR}^{[222]}(\VEC{k}_1,\VEC{k}_2) &=&  
    \left[  \ln \left( {\cal D}(\VEC{k}_3){\cal D}^{-1}(\VEC{k}_1){\cal D}^{-1}(\VEC{k}_2)  \right) \right]
    \nonumber \\
    &\times&
    B_{\rm GG}^{[112]}(\VEC{k}_1,\VEC{k}_2)  P_{\rm lin}(k_1)P_{\rm lin}(k_2)\;.
    \label{Eq:B_1loop_IR}
\end{eqnarray} 
The expressions presented here are identical to those of Eq.~(\ref{Eq:Bdecom_tree}) when expanded to the 1-loop level. Moreover, the five components of the 1-loop bispectrum cancel each other out in the IR limit:
\begin{eqnarray}
    && B_{\rm 1\mathchar`-loop,\, IR}(\VEC{k}_1,\VEC{k}_2) \nonumber \\
    &=& \left[B_{\rm GG,\, IR}^{[114]}(\VEC{k}_1,\VEC{k}_2) + B_{\rm GG,\, IR}^{\rm [123-I]}(\VEC{k}_1,\VEC{k}_2)  \right]P_{\rm lin}(k_1)P_{\rm lin}(k_2) \nonumber \\
    &+& B^{\rm [123-II]}_{\rm GM,\, IR}(\VEC{k}_1,\VEC{k}_2) P_{\rm lin}(k_1) + B^{\rm [123-III]}_{\rm MG,\, IR}(\VEC{k}_1,\VEC{k}_2) P_{\rm lin}(k_2)\nonumber \\
    &+& B_{\rm MMM,\, IR}^{[222]}(\VEC{k}_1,\VEC{k}_2) \nonumber \\
    &=& 0\;,
    \label{Eq:B1loop_IR}
\end{eqnarray}
This result indicates the IR cancellation of the bispectrum at the 1-loop level.

Assuming the absence of correlation between $\delta_{\rm (S)}$ and $\YY$ in Eq.~(\ref{Eq:delta_S_PT}), each component of the short-wavelength bispectrum at the 1-loop level is given by
\begin{eqnarray}
    B_{\rm (S)GG}^{[114]} &=& B_{\rm GG}^{[114]} - B_{\rm GG,\,IR}^{[114]}\;, \nonumber \\
    B_{\rm (S)GG}^{\rm [123-I]} &=& B_{\rm GG}^{\rm [123-I]} - B_{\rm GG,\,IR}^{\rm [123-I]}\;, \nonumber \\
    B_{\rm (S)GM}^{\rm [123-II]} &=& B_{\rm GM}^{\rm [123-II]} - B_{\rm GM,\,IR}^{\rm [123-II]}\;, \nonumber \\
    B_{\rm (S)MG}^{\rm [123-III]} &=& B_{\rm MG}^{\rm [123-III]} - B_{\rm MG,\,IR}^{\rm [123-III]}\;,\nonumber \\
    B_{\rm (S)MMM}^{[222]} &=& B_{\rm MMM}^{[222]} - B_{\rm MMM,\,IR}^{[222]}\;.
    \label{Eq:B_S}
\end{eqnarray}
In the 1-loop level, the components of the short-wavelength bispectrum are defined as the normal components subtracted by the values at the IR limit. However, it can be shown that the short-wavelength 1-loop bispectrum converges to the normal bispectrum using Eq.~(\ref{Eq:B1loop_IR}):
\begin{eqnarray}
    && B_{\rm (S)1\mathchar`-loop}(\VEC{k}_1,\VEC{k}_2) \nonumber \\
    &=& \left[B_{\rm (S)GG}^{[114]}(\VEC{k}_1,\VEC{k}_2) + B_{\rm (S)GG}^{\rm [123-I]}(\VEC{k}_1,\VEC{k}_2)  \right]P_{\rm lin}(k_1)P_{\rm lin}(k_2) \nonumber \\
    &+& B^{\rm [123-II]}_{\rm (S)GM}(\VEC{k}_1,\VEC{k}_2) P_{\rm lin}(k_1) + B^{\rm [123-III]}_{\rm (S)MG}(\VEC{k}_1,\VEC{k}_2) P_{\rm lin}(k_2)\nonumber \\
    &+& B_{\rm (S)MMM}^{[222]}(\VEC{k}_1,\VEC{k}_2) \nonumber \\
    &=& B_{\rm 1\mathchar`-loop}(\VEC{k}_1,\VEC{k}_2) \;.
    \label{Eq:B_1loop_S}
\end{eqnarray}
This result is what expected from Eq.~(\ref{Eq:IRcancel_B}), like the tree-level case in Eq.~(\ref{Eq:B_S_tree}).

Eq.~(\ref{Eq:B_1loop_IR}) is a general expression that includes the effects of reconstruction, RSD, and bias. However, it may be perceived as complex. To facilitate an intuitive understanding for the reader, we summarize the solutions for the high-$k$ limit of the 1-loop bispectrum in the simplest case, i.e., in the context of dark matter in real space before reconstruction, in Appendix~\ref{Sec:HighK}.

\subsection{IR cancellation}
\label{Sec:OneLoopIRcancel}

Consider that the bispectrum includes contributions up to the 1-loop correction in SPT, and that terms higher than the 1-loop level arise from IR effects. In this case, each component of the bispectrum, when decomposed by the $\Gamma$-expansion method, can be calculated non-perturbatively by applying Eqs.~(\ref{Eq:e_DE}), (\ref{Eq:B_S_tree}), and (\ref{Eq:B_S}) to Eq.~(\ref{Eq:IRcancel_B}):
\begin{widetext}
\begin{eqnarray}
     B_{\rm GG,\, IR}(\VEC{k}_1,\VEC{k}_2) 
    &=&{\cal D}(\VEC{k}_1){\cal D}(\VEC{k}_2){\cal D}(\VEC{k}_3) 
    \left[B_{\rm GG}^{[112]}(\VEC{k}_1,\VEC{k}_2) + B_{\rm (S)GG}^{[114]}(\VEC{k}_1,\VEC{k}_2) + B_{\rm (S)GG}^{\rm [123-I]}(\VEC{k}_1,\VEC{k}_2) \right]\;, 
    \nonumber \\
     B_{\rm GM,\, IR}(\VEC{k}_1,\VEC{k}_2) 
    &=&
    \big[{\cal D}^2(\VEC{k}_1) - {\cal D}(\VEC{k}_1){\cal D}(\VEC{k}_2){\cal D}(\VEC{k}_3) \big]
    \left[ B_{\rm GG}^{[112]}(\VEC{k}_1,\VEC{k}_2) + B_{\rm (S)GG}^{[114]}(\VEC{k}_1,\VEC{k}_2) 
    + B_{\rm (S)GG}^{\rm [123-I]}(\VEC{k}_1,\VEC{k}_2) \right]P_{\rm lin}(k_2) \nonumber \\
    &+& {\cal D}^2(\VEC{k}_1)    B_{\rm (S)GM}^{\rm [123-II]}(\VEC{k}_1,\VEC{k}_2)\;,  \nonumber \\
    B_{\rm MG,\, IR}(\VEC{k}_1,\VEC{k}_2) 
   &=&
    \big[{\cal D}^2(\VEC{k}_2) - {\cal D}(\VEC{k}_1){\cal D}(\VEC{k}_2){\cal D}(\VEC{k}_3) \big]
    \left[ B_{\rm GG}^{[112]}(\VEC{k}_1,\VEC{k}_2) + B_{\rm (S)GG}^{[114]}(\VEC{k}_1,\VEC{k}_2)
    + B_{\rm (S)GG}^{\rm [123-I]}(\VEC{k}_1,\VEC{k}_2)    \right] P_{\rm lin}(k_1)
    \nonumber \\
    &+& {\cal D}^2(\VEC{k}_2) B_{\rm (S)MG}^{\rm [123-III]}(\VEC{k}_1,\VEC{k}_2) \;, \nonumber \\
     B_{\rm MM,\, IR}(\VEC{k}_1,\VEC{k}_2) 
    &=&
    \big[ {\cal D}(\VEC{k}_1){\cal D}(\VEC{k}_2){\cal D}^{-1}(\VEC{k}_3) - {\cal D}^2(\VEC{k}_1)
         - {\cal D}^2(\VEC{k}_2) + {\cal D}(\VEC{k}_1){\cal D}(\VEC{k}_2){\cal D}(\VEC{k}_3) \big]
    \nonumber \\
    &\times&
    \left[ B_{\rm GG}^{[112]}(\VEC{k}_1,\VEC{k}_2) + B_{\rm (S)GG}^{[114]}(\VEC{k}_1,\VEC{k}_2) + B_{\rm (S)GG}^{\rm [123-I]}(\VEC{k}_1,\VEC{k}_2)  \right]
    P_{\rm lin}(k_1) P_{\rm lin}(k_2)
     \nonumber \\
   &+&
   \big[ {\cal D}(\VEC{k}_1){\cal D}(\VEC{k}_2){\cal D}^{-1}(\VEC{k}_3) - {\cal D}^2(\VEC{k}_1)\big]
   B_{\rm (S)GM}^{\rm [123-II]}(\VEC{k}_1,\VEC{k}_2)P_{\rm lin}(k_1)\nonumber \\
    &+&
   \big[ {\cal D}(\VEC{k}_1){\cal D}(\VEC{k}_2){\cal D}^{-1}(\VEC{k}_3) - {\cal D}^2(\VEC{k}_2)\big]
   B_{\rm (S)MG}^{\rm [123-III]}(\VEC{k}_1,\VEC{k}_2)P_{\rm lin}(k_2)\;, \nonumber \\
    B_{\rm MMM,\, IR}(\VEC{k}_1,\VEC{k}_2) 
    &=&
    \big[ 1 - {\cal D}(\VEC{k}_1){\cal D}(\VEC{k}_2){\cal D}^{-1}(\VEC{k}_3) \big] \nonumber \\
    &\times&
    \bigg\{\left[ B_{\rm GG}^{[112]}(\VEC{k}_1,\VEC{k}_2) +  B_{\rm (S)GG}^{[114]}(\VEC{k}_1,\VEC{k}_2) + B_{\rm (S)GG}^{\rm [123-I]}(\VEC{k}_1,\VEC{k}_2) \right] P_{\rm lin}(k_1) P_{\rm lin}(k_2) 
    \nonumber \\
    &+& B_{\rm (S)GM}^{\rm [123-II]}(\VEC{k}_1,\VEC{k}_2) P_{\rm lin}(k_1)
    + B_{\rm (S)MG}^{\rm [123-III]}(\VEC{k}_1,\VEC{k}_2) P_{\rm lin}(k_2)  \bigg\}
    + B_{\rm (S)MMM}^{[222]}(\VEC{k}_1,\VEC{k}_2) \;.
    \label{Eq:Bdecom_1loop}
\end{eqnarray}
\end{widetext}
Once again, substituting Eq.~(\ref{Eq:Bdecom_1loop}) into Eq.~(\ref{Eq:Bdecom}) leads to the 1-loop bispectrum in SPT due to the IR cancellation:
\begin{eqnarray}
    B_{\rm IR}(\VEC{k}_1,\VEC{k}_2)
    &=& B_{\rm GG,\,IR}(\VEC{k}_1,\VEC{k}_2) P_{\rm lin}(k_1)P_{\rm lin}(k_2) \nonumber \\
    &+& B_{\rm GM,\,IR}(\VEC{k}_1,\VEC{k}_2) P_{\rm lin}(k_1) \nonumber \\
    &+& B_{\rm MG,\,IR}(\VEC{k}_1,\VEC{k}_2) P_{\rm lin}(k_2)\nonumber \\
    &+& B_{\rm MM,\,IR}(\VEC{k}_1,\VEC{k}_2) \nonumber \\
    &+& B_{\rm MMM,\,IR}(\VEC{k}_1,\VEC{k}_2) \nonumber \\
    &=& B_{\rm tree}(\VEC{k}_1,\VEC{k}_2) + B_{\rm 1\mathchar`-loop}(\VEC{k}_1,\VEC{k}_2) \;.
\end{eqnarray}

\subsection{IR-resummed model}

We now proceed to derive the IR-resummed bispectrum model at the 1-loop level. To this end, analogous to the approach in Section~\ref{Sec:IR_resummed_model}, we replace the linear matter power spectra, $P_{\rm lin}$, appearing in $B_{\rm GM}$, $B_{\rm MG}$, $B_{\rm MM}$, and $B_{\rm MMM}$ in Eq.~(\ref{Eq:Bdecom_1loop}), with the no-wiggle power spectra, $P_{\rm nw}$. Note that the components of the short-wavelength bispectrum at the 1-loop level, i.e., $B_{\rm (S)GG}^{[114]}$, $B_{\rm (S)GG}^{\rm [123-I]}$, $B_{\rm (S)GM}^{\rm [123-II]}$, $B_{\rm (S)MG}^{\rm [123-III]}$, and $B_{\rm (S)MMM}^{[222]}$, are defined as the original components minus their corresponding IR limit values. Consequently, $P_{\rm lin}$ appearing in $B_{\rm GM,\,IR}^{\rm [123-II]}$, $B_{\rm MG,\,IR}^{\rm [123-III]}$, and $B_{\rm MMM,\,IR}^{[222]}$ in Eq.~(\ref{Eq:B_S}) should also be replaced with $P_{\rm nw}$. 

Finally, after replacing $P_{\rm lin}$ and $P_{\rm nw}$, substituting Eq.~(\ref{Eq:Bdecom_1loop}) into Eq.~(\ref{Eq:Bdecom}) leads to the post-reconstruction 1-loop bispectrum model with IR resummation:
\begin{widetext}
\begin{eqnarray}
    && B(\VEC{k}_1,\VEC{k}_2) \nonumber \\
    &=&
    \left[ \left[ 1 - \ln \left(  {\cal D}(\VEC{k}_1){\cal D}(\VEC{k}_2){\cal D}(\VEC{k}_3)  \right) \right]
        B_{\rm GG}^{[112]}(\VEC{k}_1,\VEC{k}_2) 
        + B_{\rm GG}^{[114]}(\VEC{k}_1,\VEC{k}_2) + B_{\rm GG}^{\rm [123-I]}(\VEC{k}_1,\VEC{k}_2)  \right]
    {\cal D}(\VEC{k}_1){\cal D}(\VEC{k}_2){\cal D}(\VEC{k}_3)P_{\rm w}(k_1)P_{\rm w}(k_2) \nonumber\\
    &+& 
    \left[ \left[ 1 - \ln {\cal D}^2(\VEC{k}_1) \right] B_{\rm GG}^{[112]}(\VEC{k}_1,\VEC{k}_2)  
             + B_{\rm GG}^{[114]}(\VEC{k}_1,\VEC{k}_2)
             + B_{\rm GG}^{\rm [123-I]}(\VEC{k}_1,\VEC{k}_2)
             + \left(B_{\rm GM}^{\rm [123-II]}(\VEC{k}_1,\VEC{k}_2) / P_{\rm nw}(k_2)\right)\right]
     {\cal D}^2(\VEC{k}_1) P_{\rm w}(k_1)P_{\rm nw}(k_2)\nonumber\\
    &+&  \left[ \left[ 1 - \ln {\cal D}^2(\VEC{k}_2) \right]
     B_{\rm GG}^{[112]}(\VEC{k}_1,\VEC{k}_2)
     + B_{\rm GG}^{[114]}(\VEC{k}_1,\VEC{k}_2)
     + B_{\rm GG}^{\rm [123-I]}(\VEC{k}_1,\VEC{k}_2)
     + \left(  B_{\rm MG}^{\rm [123-III]}(\VEC{k}_1,\VEC{k}_2) / P_{\rm nw}(k_1)\right) \right]
    {\cal D}^2(\VEC{k}_2)P_{\rm nw}(k_1)P_{\rm w}(k_2) \nonumber\\
    &+& 
    B_{\rm tree,nw}(\VEC{k}_1,\VEC{k}_2) + B_{\rm 1\mathchar`-loop, nw}(\VEC{k}_1,\VEC{k}_2)\;, 
    \label{Eq:main2}
\end{eqnarray}
\end{widetext}
where the no-wiggle versions of the 1-loop bispectrum correction is given by
\begin{eqnarray}
    && B_{\rm 1\mathchar`-loop,nw}(\VEC{k}_1,\VEC{k}_2) \nonumber \\
    &=& \left[ B_{\rm GG}^{[114]}(\VEC{k}_1,\VEC{k}_2) + B_{\rm GG}^{\rm [123-I]}(\VEC{k}_1,\VEC{k}_2) \right]P_{\rm nw}(k_1)P_{\rm nw}(k_2) \nonumber \\
    &+& B_{\rm GM}^{\rm [123-II]}(\VEC{k}_1,\VEC{k}_2) P_{\rm nw}(k_1) + B_{\rm MG}^{\rm [123-III]}(\VEC{k}_1,\VEC{k}_2) P_{\rm nw}(k_2) \nonumber \\
    &+& B_{\rm MMM}^{[222]}(\VEC{k}_1,\VEC{k}_2)\;.
\end{eqnarray}
This is the second main result of this paper. Compared to the IR-resummed model at the tree-level, as given in Eq.~(\ref{Eq:main1}), we can see that the SPT 1-loop correction terms appear in the three terms that include $P_{\rm w}$ and describe the nonlinear damping of the BAO signal. In addition, the overall shape is represented by the no-wiggle version of the SPT 1-loop bispectrum. Consequently, our model describes the non-perturbative behavior of the BAO exponential damping while incorporating 1-loop correction terms in SPT, enabling theoretical predictions of the bispectrum up to small scales. 

When perturbatively expanded, Eq.~(\ref{Eq:main2}) can be schematically expressed as the SPT 1-loop solutions with higher-order terms arising from the IR effects:
\begin{eqnarray}
    B(\VEC{k}_1,\VEC{k}_2)
    &=& B_{\rm tree}(\VEC{k}_1,\VEC{k}_2)
    + B_{\rm 1\mathchar`-loop}(\VEC{k}_1,\VEC{k}_2)\nonumber \\
    &+& \mbox{[higher order corrections]} \;.
\end{eqnarray}
This result illustrates the essence of the IR-resummed model, which accurately incorporates the SPT solution at a finite order and focuses only on IR effects for the higher-order correction terms. It is particularly noteworthy that the result in Eq.~(\ref{Eq:main2}), where the no-wiggle version of the 1-loop solution shapes the overall form of the bispectrum, is achieved by summing IR effects to an infinite order. This has a fundamentally different implication than merely truncating the SPT solution at the 1-loop level, despite the apparent similarity in form.

\subsection{Comparison with previous works}

Eq.~(\ref{Eq:main2}) represents an improvement in two aspects compared to the IR-resummed model for the pre-reconstruction bispectrum presented in \citet{Sugiyama:2020uil}. First, it includes the 1-loop correction terms in SPT, which means that it can make more accurate theoretical predictions for smaller scales. Second, it uses the post-reconstruction nonlinear kernel functions (\ref{Eq:PostReconZ}) and the post-reconstruction Gaussian smoothing factors (\ref{Eq:Sigma}), making it a model for the post-reconstruction bispectrum with IR resummation. To reiterate, the functional form of the IR-resummed model remains unchanged before and after reconstruction. Therefore, if one wants to compute the pre-reconstruction bispectrum, it is sufficient to simply replace the nonlinear kernel functions and smoothing parameters with those from before reconstruction.

Finally, it is noteworthy to mention that \citet{Ivanov:2018gjr} has already presented an IR-resummed model for the pre-reconstruction bispectrum, including the SPT 1-loop correction terms, RSD effects, and bias effects, through a different approach from ours, the time-sliced perturbation theory~\cite{Blas:2016sfa}. The form of the bispectrum in the \citet{Ivanov:2018gjr}'s model is schematically given as follows (for details, see Eq.~(7.19) and Appendix D in~\cite{Ivanov:2018gjr}):
\begin{eqnarray}
    B&=&  B_{\rm tree}\left[ P_{\rm nw} + \left( 1-\ln {\cal D}^2 \right){\cal D}^2 P_{\rm w} \right]
    \nonumber \\
    &+& B_{\rm 1\mathchar`-loop}\left[ P_{\rm nw} + {\cal D}^2 P_{\rm w} \right] \;.
\end{eqnarray}
This expression indicates that the linear power spectrum appearing within the bispectrum is replaced with $P_{\rm nw} + {\cal D}^2 P_{\rm w}$. In the tree-level bispectrum, a factor of $(1-\ln {\cal D}^2)$ appears to avoid double counting of nonlinear contributions. While our result in Eq.~(\ref{Eq:main2}) align closely with their model, a slight difference arises in the exponential damping functions form at the order of $(P_{\rm w})^2$. Specifically, the \citet{Ivanov:2018gjr}'s model includes ${\cal D}^2(\VEC{k}_1) {\cal D}^2(\VEC{k}_2) P_{\rm w}(k_1)P_{\rm w}(k_2)$, whereas our result is ${\cal D}(\VEC{k}_1){\cal D}(\VEC{k}_2){\cal D}(\VEC{k}_3)P_{\rm w}(k_1)P_{\rm w}(k_2)$. For a more intuitive comparison, when considered in real space as in Appendix~\ref{Sec:HighK}, the \citet{Ivanov:2018gjr}'s model includes
\begin{eqnarray}
    && {\cal D}^2(\VEC{k}_1) {\cal D}^2(\VEC{k}_2) P_{\rm w}(k_1)P_{\rm w}(k_2) \nonumber \\
    &=& 
    e^{-k_1^2\sigma_{\perp}^2}e^{-k_2^2\sigma_{\perp}^2} P_{\rm w}(k_1)P_{\rm w}(k_2), 
\end{eqnarray}
while our model is
\begin{eqnarray}
    &&{\cal D}(\VEC{k}_1) {\cal D}(\VEC{k}_2){\cal D}(\VEC{k}_3)  P_{\rm w}(k_1)P_{\rm w}(k_2) \nonumber \\
    &=& 
    e^{-k_1^2\sigma_{\perp}^2}e^{-k_2^2\sigma_{\perp}^2}
    e^{-(\VEC{k}_1\cdot\VEC{k}_2)\sigma_{\perp}^2} P_{\rm w}(k_1)P_{\rm w}(k_2) \;,
\end{eqnarray}
differing by a factor of $e^{-(\VEC{k}_1\cdot\VEC{k}_2)\sigma_{\perp}^2}$. We conjecture that this discrepancy arises due to the approximation employed by \citet{Ivanov:2018gjr}, who ignored terms of ${\cal O}(P_{\rm w}^2)$ during their derivations. However, a detailed mathematical comparison or a discussion on numerical impacts exceeds the scope of this paper and is left for future work.

\section{Conclusions}
\label{Sec:Conclusions}

In this paper, we have developed a bispectrum model that incorporates infrared (IR) effects into post-reconstruction density fluctuations, including 1-loop corrections within SPT. The main results are presented in Eqs.~(\ref{Eq:main1}) and (\ref{Eq:main2}). This model effectively captures the nonlinear dynamics of the BAO signal within the post-reconstruction bispectrum. Moreover, it facilitates analysis on smaller scales via the no-wiggle version of the 1-loop solution. Note that the no-wiggle version of the SPT solution, which shapes the overall bispectrum form, can only be derived by properly summing up the IR effects to an infinite order. This approach fundamentally differs from merely truncating the SPT solution at the 1-loop level. By setting the smoothing parameter $R_{\rm s}$, required for reconstruction in Eq.~(\ref{Eq:S}), to approach infinity $(R_{\rm s})\to \infty$, our model encompasses the pre-reconstruction case.

\citet{Sugiyama:2024eye} demonstrated that the IR effect can be treated consistently before and after reconstruction. Specifically, the IR effect can be characterized as a coordinate transformation through the displacement vector in short-wavelength density fluctuations in both cases. This fact suggests that the resummation method for addressing the IR effect in the pre-reconstruction tree-level bispectrum, as developed by \citet{Sugiyama:2020uil}, can be directly applied to the post-reconstruction bispectrum. In this paper, we build upon the framework established by \citet{Sugiyama:2020uil} by incorporating the 1-loop correction terms. Consequently, our work extends previous research in two aspects: firstly, by incorporating the SPT 1-loop correction terms, and secondly, by considering the post-reconstruction scenario.

The characteristics of the IR-resummed bispectrum model handle terms that include infinite-order mode-coupling integrals. Through these mode-coupling terms, the bispectrum is decomposed into five components, as demonstrated in Eq.~(\ref{Eq:Bdecom}). We calculate the non-perturbative effects arising from the IR effects for each component. To streamline the overall computation and minimize complex calculations involving kernel functions, we introduce the functions ${\cal D}$ and ${\cal E}$ in Eq.~(\ref{Eq:D_E}). In the IR limit, statistical translational symmetry requires the cancellation of all IR effects. Nonetheless, by considering an incompleteness in the IR cancellation, namely the smoothing effect of the BAO signal in the mode-coupling integrals, we are enable to formulate a model that describes the nonlinear damping of the BAO signal while incorporating the 1-loop bispectrum correction terms.

The bispectrum model proposed in this paper encompasses 1-loop correction terms, bias effects, and RSD effects, making it a suitable candidate for cosmological analysis of the post-reconstruction bispectrum.

\begin{acknowledgments}
NS acknowledges financial support from JSPS KAKENHI Grant Number 19K14703. 
NS thanks Shi-Fan Stephen Chen for useful comments and discussion.
\end{acknowledgments}

\appendix

\section{Post-reconstruction kernel functions}
\label{Sec:Kernel}

The $n$th-order pre-reconstruction galaxy density fluctuation in Fourier space is expressed as
\begin{eqnarray}
    \hspace{-0.3cm}
    \widetilde{\delta}_{\rm g}^{\,[n]}(\VEC{k})
    &=& \int \frac{d^3p_1}{(2\pi)^3}\cdots\int \frac{d^3p_n}{(2\pi)^3}
    (2\pi)^3\delta_{\rm D}\left( \VEC{k} - \VEC{p}_{[1,n]} \right) \nonumber \\
    &\times& Z^{[n]}(\VEC{p}_1,\cdots,\VEC{p}_n) \widetilde{\delta}_{\rm lin}(\VEC{p}_1)\cdots\widetilde{\delta}_{\rm lin}(\VEC{p}_n)\;,
    \label{Eq:Z}
\end{eqnarray}
where the functions $Z^{[n\geq2]}$ characterize the nonlinear effects on the galaxy density fluctuation, including the RSD and bias effects. As in the main text, for simplicity of notation, the line-of-sight dependence $\hat{n}$ due to the RSD effect is omitted. This appendix shows the relation between $Z^{[n]}$ in Eq.~(\ref{Eq:Z}) and $Z^{[n]}_{\rm rec}$ in Eq.~(\ref{Eq:Zrec}) up to the forth order.

The $n$th-order displacement vector for reconstruction in Fourier space is given by
\begin{eqnarray}
    \widetilde{\VEC{s}}^{\,[n]}(\VEC{k})
    = \left( \frac{i\VEC{k}}{k^2} \right)
    \left(  - \frac{W_{\rm G}(pR_{\rm s})}{b_{1, \rm fid}} \right)
    \widetilde{\delta}_{\rm g}^{\,[n]}(\VEC{k})\;.
    \label{Eq:S_F}
\end{eqnarray}
Furthermore, the Fourier transform of the post-reconstruction density fluctuation in Eq.~(\ref{Eq:delta_rec}) is given by
\begin{eqnarray}
    \widetilde{\delta}_{\rm rec}(\VEC{k},\hat{n})
    = \int d^3x' e^{-i\VEC{k}\cdot\VEC{x}'}
    e^{-i\VEC{k}\cdot\VEC{s}(\VEC{x}',\hat{n})}
    \delta_{\rm g}(\VEC{x}',\hat{n}) \;.
    \label{Eq:delta_rec_Fourier}
\end{eqnarray}
By substituting Eqs.~(\ref{Eq:Z}) and (\ref{Eq:S_F}) into Eq.~(\ref{Eq:delta_rec_Fourier}), we obtain
\begin{widetext}
\begin{eqnarray}
    Z_{\rm rec}^{[1]}(\VEC{p}_1) &=& Z^{[1]}(\VEC{p}_1)\;, \nonumber \\
    Z_{\rm rec}^{[2]}(\VEC{p}_1,\VEC{p}_2) &=& Z^{[2]}(\VEC{p}_1,\VEC{p}_2)
    + \frac{1}{2}\Bigg\{ \left( \frac{\VEC{k}\cdot\VEC{p}_1}{p_1^2} \right)
    \left(  - \frac{W_{\rm G}(p_1R_{\rm s})}{b_{1, \rm fid}} \right) 
    + \left( \frac{\VEC{k}\cdot\VEC{p}_2}{p_2^2} \right)
    \left(  - \frac{W_{\rm G}(p_2R_{\rm s})}{b_{1, \rm fid}} \right) \Bigg\}
    Z^{[1]}(\VEC{p}_1)
    Z^{[1]}(\VEC{p}_2)\;, \nonumber \\
    Z_{\rm rec}^{[3]}(\VEC{p}_1,\VEC{p}_2,\VEC{p}_3) &=& Z^{[3]}(\VEC{p}_1,\VEC{p}_2,\VEC{p}_3)
    \nonumber \\
    &+& \frac{1}{3} \Bigg\{\left( \frac{\VEC{k}\cdot(\VEC{p}_1+\VEC{p}_2)}{|\VEC{p}_1+\VEC{p}_2|^2} \right)
    \left(  - \frac{W_{\rm G}(|\VEC{p}_1+\VEC{p}_2|R_{\rm s})}{b_{1, \rm fid}} \right) 
    Z^{[2]}(\VEC{p}_1,\VEC{p}_2)Z^{[1]}(\VEC{p}_3) + \mbox{2 perms.}\Bigg\}
    \nonumber \\
    &+& \frac{1}{3} \Bigg\{\left( \frac{\VEC{k}\cdot\VEC{p}_1}{p_1^2} \right)
    \left(  - \frac{W_{\rm G}(p_1R_{\rm s})}{b_{1, \rm fid}} \right) 
    Z^{[1]}(\VEC{p}_1)Z^{[2]}(\VEC{p}_2,\VEC{p}_3) + \mbox{2 perms.}\Bigg\}
    \nonumber \\
    &+& \frac{1}{3!}\Bigg\{ \left( \frac{\VEC{k}\cdot\VEC{p}_1}{p_1^2} \right)
    \left(  - \frac{W_{\rm G}(p_1R_{\rm s})}{b_{1, \rm fid}} \right) 
    \left( \frac{\VEC{k}\cdot\VEC{p}_2}{p_2^2} \right)
    \left(  - \frac{W_{\rm G}(p_2R_{\rm s})}{b_{1, \rm fid}} \right) 
    + \mbox{2 perms.} \Bigg\}
    Z^{[1]}(\VEC{p}_1)
    Z^{[1]}(\VEC{p}_2)
    Z^{[1]}(\VEC{p}_3) \;, \nonumber \\
    Z_{\rm rec}^{[4]}(\VEC{p}_1,\VEC{p}_2,\VEC{p}_3,\VEC{p}_4) &=& Z^{[4]}(\VEC{p}_1,\VEC{p}_2,\VEC{p}_3,\VEC{p}_4)
    \nonumber \\
    &+&
    \frac{1}{4}
    \Bigg\{\left( \frac{\VEC{k}\cdot(\VEC{p}_1+\VEC{p}_2+\VEC{p}_3)}{|\VEC{p}_1+\VEC{p}_2+\VEC{p}_3|^2} \right)
    \left(  - \frac{W_{\rm G}(|\VEC{p}_1+\VEC{p}_2+\VEC{p}_3|R_{\rm s})}{b_{1, \rm fid}} \right) 
    Z^{[3]}(\VEC{p}_1,\VEC{p}_2,\VEC{p}_3)Z^{[1]}(\VEC{p}_4) + \mbox{4 perms.}\Bigg\}
    \nonumber \\
    &+& 
    \frac{1}{4}
    \Bigg\{\left( \frac{\VEC{k}\cdot\VEC{p}_1}{p_1^2} \right)
    \left(  - \frac{W_{\rm G}(p_1R_{\rm s})}{b_{1, \rm fid}} \right) 
    Z^{[1]}(\VEC{p}_1)Z^{[3]}(\VEC{p}_2,\VEC{p}_3,\VEC{p}_4) + \mbox{4 perms.}\Bigg\}
    \nonumber \\
    &+&
    \frac{1}{6}\Bigg\{\left( \frac{\VEC{k}\cdot(\VEC{p}_1+\VEC{p}_2)}{|\VEC{p}_1+\VEC{p}_2|^2} \right)
    \left(  - \frac{W_{\rm G}(|\VEC{p}_1+\VEC{p}_2|R_{\rm s})}{b_{1, \rm fid}} \right) 
    Z^{[2]}(\VEC{p}_1,\VEC{p}_2)Z^{[2]}(\VEC{p}_3,\VEC{p}_4) + \mbox{5 perms.}\Bigg\}
    \nonumber \\
    &+&
   \frac{1}{12}\Bigg\{  \left( \frac{\VEC{k}\cdot(\VEC{p}_1+\VEC{p}_2)}{|\VEC{p}_1+\VEC{p}_2|^2} \right)
   \left(  - \frac{W_{\rm G}(|\VEC{p}_1+\VEC{p}_2|R_{\rm s})}{b_{1, \rm fid}} \right) 
    \left( \frac{\VEC{k}\cdot\VEC{p}_3}{p_3^2} \right)
    \left(  - \frac{W_{\rm G}(p_3R_{\rm s})}{b_{1, \rm fid}} \right) \nonumber \\
    && \hspace{1cm} \times
    Z^{[2]}(\VEC{p}_1,\VEC{p}_2) 
    Z^{[1]}(\VEC{p}_3)
    Z^{[1]}(\VEC{p}_4)
    + \mbox{11 perms.}
    \Bigg\}
    \nonumber \\
   &+& \frac{1}{12}
   \Bigg\{
   \left( \frac{\VEC{k}\cdot\VEC{p}_1}{p_1^2} \right)
   \left(  - \frac{W_{\rm G}(p_1R_{\rm s})}{b_{1, \rm fid}} \right) 
   \left( \frac{\VEC{k}\cdot\VEC{p}_2}{p_2^2} \right)
   \left(  - \frac{W_{\rm G}(p_2R_{\rm s})}{b_{1, \rm fid}} \right) 
   Z^{[1]}(\VEC{p}_1) Z^{[1]}(\VEC{p}_2) Z^{[2]}(\VEC{p}_3,\VEC{p}_4)
   + \mbox{5 perms.} \Bigg\} \nonumber \\
    &+&
    \frac{1}{4!}
    \Bigg\{\left( \frac{\VEC{k}\cdot\VEC{p}_1}{p_1^2} \right)
    \left(  - \frac{W_{\rm G}(p_1R_{\rm s})}{b_{1, \rm fid}} \right) 
    \left( \frac{\VEC{k}\cdot\VEC{p}_2}{p_2^2} \right)
    \left(  - \frac{W_{\rm G}(p_2R_{\rm s})}{b_{1, \rm fid}} \right) 
    \left( \frac{\VEC{k}\cdot\VEC{p}_3}{p_3^2} \right)
    \left(  - \frac{W_{\rm G}(p_3R_{\rm s})}{b_{1, \rm fid}} \right) 
    + \mbox{3 perms.}
    \Bigg\}
    \nonumber \\
    && \hspace{1cm} \times
    Z^{[1]}(\VEC{p}_1) Z^{[1]}(\VEC{p}_2) Z^{[1]}(\VEC{p}_3) Z^{[1]}(\VEC{p}_4) \;,
    \label{Eq:PostReconZ}
\end{eqnarray}
\end{widetext}
where $\VEC{k}=\VEC{p}_1+\cdots+\VEC{p}_n$ for $Z^{[n]}$. In the IR limit, these kernel functions become
\begin{widetext}
\begin{eqnarray}
    Z^{[2]}_{\rm rec}(\VEC{k},\VEC{p}) &\xrightarrow[p\to0]{}& 
   \frac{1}{2}     \left\{ \left( \frac{\VEC{k}\cdot\MAT{R}^{[1]}\cdot\VEC{p}}{p^2} \right) 
    + \left( \frac{\VEC{k}\cdot\VEC{p}}{p^2} \right) \left( -\frac{W_{\rm G}(pR_{\rm s})}{b_{\rm 1,fid}} \right) Z^{[1]}(\VEC{p}) \right\}
    Z^{[1]}(\VEC{k})\;,\nonumber \\
    Z^{[3]}_{\rm rec}(\VEC{k}_1,\VEC{p},-\VEC{p}) &\xrightarrow[p\to0]{}& 
   -\frac{1}{3!}
    \left\{ \left( \frac{\VEC{k}\cdot\MAT{R}^{[1]}\cdot\VEC{p}}{p^2} \right) 
    + \left( \frac{\VEC{k}\cdot\VEC{p}}{p^2} \right) 
    \left( -\frac{W_{\rm G}(pR_{\rm s})}{b_{\rm 1,fid}} \right) Z^{[1]}(\VEC{p}) \right\}^2 
    Z^{[1]}(\VEC{k})\;, \nonumber \\
    Z^{[3]}_{\rm rec}(\VEC{k}_1,\VEC{k}_2,\VEC{p}) &\xrightarrow[p\to0]{}& 
    \frac{1}{3}
    \left\{ \left( \frac{(\VEC{k}_1+\VEC{k}_2)\cdot\MAT{R}^{[1]}\cdot\VEC{p}}{p^2} \right) 
    + \left( \frac{(\VEC{k}_1+\VEC{k}_2)\cdot\VEC{p}}{p^2} \right) \left( -\frac{W_{\rm G}(pR_{\rm s})}{b_{\rm 1,fid}} \right) Z^{[1]}(\VEC{p}) \right\}
    Z_{\rm rec}^{[2]}(\VEC{k}_1,\VEC{k}_2)
    \;,\nonumber \\
    Z^{[4]}_{\rm rec}(\VEC{k}_1,\VEC{k}_2,\VEC{p},-\VEC{p}) &\xrightarrow[p\to0]{}& 
   -\frac{1}{12}
   \left\{ \left( \frac{(\VEC{k}_1+\VEC{k}_2)\cdot\MAT{R}^{[1]}\cdot\VEC{p}}{p^2} \right) 
   + \left( \frac{(\VEC{k}_1+\VEC{k}_2)\cdot\VEC{p}}{p^2} \right) 
    \left( -\frac{W_{\rm G}(pR_{\rm s})}{b_{\rm 1,fid}} \right) Z^{[1]}(\VEC{p}) \right\}^2 
    Z_{\rm rec}^{[2]}(\VEC{k}_1,\VEC{k}_2) \;,
    \label{Eq:Z_rec_IR}
\end{eqnarray}
\end{widetext}
where $R^{[1]}_{ij} = \delta_{ij} + f \hat{n}_i\hat{n}_j$ given in Eq.~(\ref{Eq:R}). In the derivation of the above relations, we used
\begin{widetext}
\begin{eqnarray}
    Z^{[2]}(\VEC{k},\VEC{p}) &\xrightarrow[p\to0]{}& 
   \frac{1}{2} \left( \frac{\VEC{k}\cdot\MAT{R}^{[1]}\cdot\VEC{p}}{p^2} \right)     Z^{[1]}(\VEC{k})\;,\nonumber \\
    Z^{[3]}(\VEC{k}_1,\VEC{p},-\VEC{p}) &\xrightarrow[p\to0]{}& 
   -\frac{1}{3!}
    \left\{ \left( \frac{\VEC{k}\cdot\MAT{R}^{[1]}\cdot\VEC{p}}{p^2} \right)  \right\}^2 
    Z^{[1]}(\VEC{k})\;, \nonumber \\
    Z^{[3]}(\VEC{k}_1,\VEC{k}_2,\VEC{p}) &\xrightarrow[p\to0]{}& 
    \frac{1}{3}
    \left\{ \left( \frac{(\VEC{k}_1+\VEC{k}_2)\cdot\MAT{R}^{[1]}\cdot\VEC{p}}{p^2} \right) \right\}
    Z^{[2]}(\VEC{k}_1,\VEC{k}_2)
    \;,\nonumber \\
    Z^{[4]}(\VEC{k}_1,\VEC{k}_2,\VEC{p},-\VEC{p}) &\xrightarrow[p\to0]{}& 
   -\frac{1}{12}
   \left\{ \left( \frac{(\VEC{k}_1+\VEC{k}_2)\cdot\MAT{R}^{[1]}\cdot\VEC{p}}{p^2} \right)  \right\}^2 
   Z^{[2]}(\VEC{k}_1,\VEC{k}_2) \;.
    \label{Eq:Z_IR}
\end{eqnarray}
\end{widetext}

\section{$\Gamma$-expansion}
\label{Sec:Gamma}

Using the $\Gamma$-expansion scheme, we show specific forms of $B_{\rm GG}$, $B_{\rm GM}$, $B_{\rm MG}$, $B_{\rm MM}$, and $B_{\rm MMM}$ in Eq.~(\ref{Eq:Bdecom}) using the nonlinear kernel functions $Z^{[n]}$. The results obtained here are also valid using the post-reconstruction nonlinear kernel functions $Z^{[n]}_{\rm rec}$.

The $r$th order coefficient of the $\Gamma$-expansion is expressed as~\cite{Sugiyama:2012pc,Sugiyama:2013pwa}
\begin{eqnarray}
	&& \Gamma^{(r)}(\VEC{p}_1,\dots,\VEC{p}_r) \nonumber \\
	&=&
	\frac{1}{r!} \sum_{s=0}^{\infty}\frac{(r+2s)!}{2^s s!}
	\prod_{i=1}^s \int \frac{d^3 q_i}{(2\pi)^3}\, P_{\rm lin}(q_i) \nonumber \\
	&\times&
    Z^{[r+2s]}(\VEC{p}_1,\dots,\VEC{p}_r,
	\VEC{q}_1,-\VEC{q}_1,\dots,\VEC{q}_s,-\VEC{q}_s)\;,
	\label{Eq:Gamma_Zn}
\end{eqnarray}
where the superscript $(r)$ means the $r$th order of the $\Gamma$-expansion. Then, we obtain~\cite{Sugiyama:2020uil}
\begin{widetext}
\begin{eqnarray}
	B_{\rm GG}(\VEC{k}_1,\VEC{k}_2) &=& 2\, \Gamma^{(2)}\left( \VEC{k}_1, \VEC{k}_2 \right)
	\Gamma^{(1)}\left( \VEC{k}_1\right) \Gamma^{(1)}\left( \VEC{k}_2 \right)\;, \nonumber \\
    B_{\rm GM}(\VEC{k}_1,\VEC{k}_2) 
	&=&
    \Gamma^{(1)}(\VEC{k}_1)\sum_{r=2}^{\infty}(r+1)! 
	\prod_{i=1}^{r}\int \frac{d^3p_i}{(2\pi)^3}P_{\rm lin}(p_i) 
    (2\pi)^3\delta_{\rm D}\big( \VEC{k}_2 - \VEC{p}_{[1,r]}  \big) 
	\Gamma^{(r)}(\VEC{p}_1,\dots,\VEC{p}_r) 
	\Gamma^{(r+1)}(\VEC{p}_1,\dots,\VEC{p}_r,\VEC{k}_1)\;, \nonumber \\
	B_{\rm MG}(\VEC{k}_1,\VEC{k}_2) 
	&=& 
	\Gamma^{(1)}(\VEC{k}_2)\sum_{r=2}^{\infty}(r+1)! 
	\prod_{i=1}^{r}\int \frac{d^3p_i}{(2\pi)^3}P_{\rm lin}(p_i) 
    (2\pi)^3\delta_{\rm D}\big( \VEC{k}_1 - \VEC{p}_{[1,r]}  \big) 
	\Gamma^{(r)}(\VEC{p}_1,\dots,\VEC{p}_r) 
	\Gamma^{(r+1)}(\VEC{p}_1,\dots,\VEC{p}_r,\VEC{k}_2) \;, \nonumber \\
	B_{\rm MM}(\VEC{k}_1,\VEC{k}_2) 
	&=& 
	\sum_{r=2}^{\infty}\sum_{s=2}^{\infty}(r+s)!  
    \prod_{i=1}^{r}\int \frac{d^3p_i}{(2\pi)^3}P_{\rm lin}(p_i) 
	\prod_{j=1}^{s}\int \frac{d^3p'_j}{(2\pi)^3}P_{\rm lin}(p'_j) 
    (2\pi)^3\delta_{\rm D}\big( \VEC{k}_1 - \VEC{p}_{[1,r]}  \big)  
    (2\pi)^3\delta_{\rm D}\big( \VEC{k}_2 - \VEC{p}'_{[1,s]} \big) \nonumber \\
    &\times&
	\Gamma^{(r)}(\VEC{p}_1,\dots,\VEC{p}_r)\Gamma^{(s)}(\VEC{p}'_1,\dots,\VEC{p}'_s) 
	\Gamma^{(r+s)}(\VEC{p}_1,\dots,\VEC{p}_r,\VEC{p}'_1,\dots,\VEC{p}'_s)\;, \nonumber \\
	B_{\rm MMM}(\VEC{k}_1,\VEC{k}_2) 
    &=& \frac{1}{3}
	\sum_{r=1}^{\infty}\sum_{s=1}^{\infty}\sum_{t=1}^{\infty}
	\frac{(r+t)!(r+s)!(s+t)!}{r!s!t!} 
    \prod_{i=1}^{r}\int \frac{d^3p_i}{(2\pi)^3}P_{\rm lin}(p_i) 
	\prod_{j=1}^{s}\int \frac{d^3p'_j}{(2\pi)^3} P_{\rm lin}(p'_j)
	\prod_{k=1}^{t}\int \frac{d^3p''_k}{(2\pi)^3} P_{\rm lin}(p''_k) \nonumber \\
	&\times&
    (2\pi)^3\delta_{\rm D}\big( \VEC{k}_1 - \VEC{p}_{[1,r]} - \VEC{p}''_{[1,t]} \big)  
    (2\pi)^3\delta_{\rm D}\big( \VEC{k}_2 + \VEC{p}_{[1,r]} + \VEC{p}'_{[1,s]}\big) \nonumber \\
	&\times&
	\Gamma^{(r+t)}(\VEC{p}_1,\dots,\VEC{p}_r,\VEC{p}''_1,\dots,\VEC{p}''_t) 
    \Gamma^{(r+s)}(-\VEC{p}_1,\dots,-\VEC{p}_r,-\VEC{p}'_1,\dots,-\VEC{p}'_s) \nonumber \\
    &\times&
    \Gamma^{(s+t)}(-\VEC{p}'_1,\dots,-\VEC{p}'_s, \VEC{p}''_1,\dots,\VEC{p}''_t)\;,
\end{eqnarray}
\end{widetext}
where $\VEC{p}_{[1,n]} = \VEC{p}_1+\cdots+\VEC{p}_n$ for any integer $n$ with $n\geq1$.

\section{Smoothing factors}
\label{Sec:Gaussian}

This appendix gives the specific forms of the smoothing factors $\sigma^2_{\perp}$ and $\sigma^2_{\parallel}$ that characterize the Gaussian damping function ${\cal D}(\VEC{k})$ defined in Eq.~(\ref{Eq:Damping}). 

The first-order of the post-reconstruction displacement vector is given by
\begin{eqnarray}
    \YY_{\rm rec}^{[1]} &=& i\,\int \frac{d^3p}{(2\pi)^3} 
     \Bigg\{ \left( \frac{\MAT{R}^{[1]}\cdot\VEC{p}}{p^2} \right) \nonumber \\
     && + \left( \frac{\VEC{p}}{p^2} \right) \left( -\frac{W_{\rm G}(pR_{\rm s})}{b_{\rm 1,fid}} \right) Z^{[1]}(\VEC{p}) \Bigg\} \delta_{\rm lin}(\VEC{p})\;.
     \label{Eq:YY1}
\end{eqnarray}
By substituting Eq.~(\ref{Eq:YY1}) into Eq.~(\ref{Eq:D_E}), the radial and transverse components of the smoothing factors are calculated as~\cite{Sugiyama:2024eye}
\begin{eqnarray}
    \sigma_{\perp}^2 &=&  \sigma_{\rm pp\,,\perp}^2 + \sigma_{\rm ps\,,\perp}^2 + \sigma_{\rm ss\,,\perp}^2
    \nonumber \\
    \sigma_{\parallel}^2 &=& \sigma_{\rm pp\,,\parallel}^2 
    + \sigma_{\rm ps\,,\parallel}^2 + \sigma_{\rm ss\,,\parallel}^2\;,
    \label{Eq:Sigma}
\end{eqnarray}
and
\begin{widetext}
    \begin{eqnarray}
    \sigma^2_{\rm pp,\perp} &=&
    \frac{1}{3}\int \frac{dp}{2\pi^2} P_{\rm lin}(p), \nonumber \\
    \sigma^2_{\rm pp,\parallel} &=& (1+f)^2\, \sigma^2_{\rm pp, \perp}, \nonumber \\
        \sigma^2_{\rm ps, \perp}
        &=& \frac{1}{3} \int_{k_{\rm min}}^{k_{\rm max}} \frac{dp}{2\pi^2} \left( -\frac{W_{\rm G}(pR_{\rm s})}{b_{1,\rm fid}} \right)  
        \left[ 2\left( b_1+\frac{f}{5} \right)    \right]P_{\rm lin}(p)\;, \nonumber \\
        \sigma^2_{\rm ps, \parallel}
        &=& \frac{1}{3} \int_{k_{\rm min}}^{k_{\rm max}} \frac{dp}{2\pi^2} \left( -\frac{W_{\rm G}(pR_{\rm s})}{b_{1,\rm fid}} \right)
        \Bigg[ 2\left( 1+f \right)\left(  b_1+\frac{3}{5}f\right)
              \Bigg]P_{\rm lin}(p)\;, \nonumber \\
        \sigma^2_{\rm ss, \perp}
        &=& \frac{1}{3} \int_{k_{\rm min}}^{k_{\rm max}} \frac{dp}{2\pi^2} \left( -\frac{W_{\rm G}(pR_{\rm s})}{b_{1,\rm fid}} \right)^2
        \left[ \left( b_1^2+\frac{2}{5}b_1f + \frac{3}{35}f^2 \right) P_{\rm lin}(p) + \frac{1}{\bar{n}}   \right]\;, \nonumber \\
        \sigma^2_{\rm ss, \parallel}
        &=& \frac{1}{3} \int_{k_{\rm min}}^{k_{\rm max}} \frac{dp}{2\pi^2} \left( -\frac{W_{\rm G}(pR_{\rm s})}{b_{1,\rm fid}} \right)^2 
        \Bigg[ \left( b_1^2 + \frac{42}{35}b_1f +\frac{3}{7}f^2  \right) P_{\rm lin}(p) + \frac{1}{\bar{n}}
        \Bigg]\;.
        \label{Eq:ps_ss}
    \end{eqnarray}
\end{widetext}
Here, the shot-noise term $1/\bar{n}$ with $\bar{n}$ being the mean number density appears in $\sigma_{\rm ss,\perp}^2$ and $\sigma_{\rm ss,\parallel}^2$ due to the discreteness effect in the displacement vector for reconstruction~\cite{White:2010qd}.

\section{The high-$k$ limit solutions of the matter bispectrum in real space}
\label{Sec:HighK}

The second-order density fluctuation for dark matter in real space is expressed as
\begin{eqnarray}
    \widetilde{\delta}^{\,[2]}(\VEC{k}) 
    &=& \int \frac{d^3p_1}{(2\pi)^3}\int \frac{d^3p_2}{(2\pi)^3}
    (2\pi)^3\delta_{\rm D}\left( \VEC{k}-\VEC{p}_1-\VEC{p}_2 \right) \nonumber \\
    &\times&
    F^{[2]}(\VEC{p}_1,\VEC{p}_2) \widetilde{\delta}_{\rm lin}(\VEC{p}_1)\widetilde{\delta}_{\rm lin}(\VEC{p}_2) \;,
\end{eqnarray}
where
\begin{eqnarray}
    \hspace{-0.7cm}
    && F^{[2]}(\VEC{p}_1,\VEC{p}_2) \nonumber \\
    \hspace{-0.7cm}
    &=& \frac{17}{21} + \frac{1}{2} \left( \hat{p}_1\cdot\hat{p}_2 \right) \left( \frac{p_1}{p_2} + \frac{p_2}{p_1} \right)
    + \frac{2}{7} \left( \left( \hat{p}_1\cdot\hat{p}_2 \right)^2 - \frac{1}{3} \right) \;.
\end{eqnarray}
The tree-level bispectrum is then given by
\begin{eqnarray}
    B_{\rm tree}(\VEC{k}_1,\VEC{k}_2) = 2\, F^{[2]}(\VEC{k}_1,\VEC{k}_2) P_{\rm lin}(k_1)P_{\rm lin}(k_2) \;.
\end{eqnarray}
From Eq.~(\ref{Eq:ps_ss}), the smoothing factor is calculated as
\begin{eqnarray}
    \sigma_{\perp}^2 = \frac{1}{3}\int \frac{dp}{2\pi^2}P_{\rm lin}(p) \;.
\end{eqnarray}
Consequently, Eq.~(\ref{Eq:B_1loop_IR}) can be rewritten as
\begin{widetext}
\begin{eqnarray}
    B_{\rm GG,\,IR}^{[114]}(\VEC{k}_1,\VEC{k}_2) &=&  
    \left[ - \frac{1}{2}k_1^2 \sigma^2_{\perp} - \frac{1}{2}k_1^2 \sigma^2_{\perp} - (\VEC{k}_1\cdot\VEC{k}_2) \sigma^2_{\perp}  \right]
    \left[ 2 F_2(\VEC{k}_,\VEC{k}_2) \right] \;, \nonumber \\
    B_{\rm GG,\,IR}^{\rm [123-I]}(\VEC{k}_1,\VEC{k}_2) 
    &=&  
    \left[ - \frac{1}{2}k_1^2 \sigma^2_{\perp} - \frac{1}{2}k_1^2 \sigma^2_{\perp} \right]\left[ 2 F_2(\VEC{k}_,\VEC{k}_2) \right] \;, \nonumber \\
    B_{\rm GM,\,IR}^{\rm [123-II]}(\VEC{k}_1,\VEC{k}_2) &=&  
    \left[ k_2^2 \sigma^2_{\perp} + (\VEC{k}_1\cdot\VEC{k}_2)  \sigma^2_{\perp}  \right] 
    \left[ 2 F_2(\VEC{k}_1,\VEC{k}_2) \right]  P_{\rm lin}(k_2) \;, \nonumber \\
    B_{\rm MG,\,IR}^{\rm [123-III]}(\VEC{k}_1,\VEC{k}_2) &=&  
    \left[ k_1^2 \sigma^2_{\perp} + (\VEC{k}_1\cdot\VEC{k}_2)  \sigma^2_{\perp}  \right]
    \left[ 2 F_2(\VEC{k}_1,\VEC{k}_2) \right]  P_{\rm lin}(k_1) \;, \nonumber \\
    B_{\rm MMM,\,IR}^{[222]}(\VEC{k}_1,\VEC{k}_2) &=&  
    \left[ - \left( \VEC{k}_1\cdot\VEC{k}_2 \right)\sigma^2_{\perp}\right] \left[ 2 F_2(\VEC{k}_1,\VEC{k}_2) \right]  P_{\rm lin}(k_1)P_{\rm lin}(k_2)\;.
\end{eqnarray} 
\end{widetext}
From these equations, we can see that the high-$k$ solution of the 1-loop bispectrum includes not only the squares of $k_1$ and $k_2$ but also the product of $\VEC{k}_1$ and $\VEC{k}_2$.

%
%
%
\bibliography{ms}

\end{document}